# Instability Mechanisms and Intermittency Distribution in Adverse Pressure Gradient Attached and Separated Boundary Layers

A. Samson[1], Karthik Naicker[2], and Sourabh S. Diwan *

Department of Aerospace Engineering, Indian Institute of Science, Bengaluru, India – 560012

**ABSTRACT**

Direct numerical simulation has been carried out on one attached and two separated boundary layer flows (involving small and large separation) under the influence of an adverse pressure gradient. A unified picture of the pre-transitional boundary layer for the three cases has been provided that reveals a "mixed-mode" instability, involving contribution from instability waves and streamwise streaks. A time-frequency analysis of the transitional velocity signals has been performed which shows that as the Reynolds number decreases, the character of the time traces evolves continuously from a "spotty" behaviour (exhibiting distinct turbulent spots) for the attached case to a "non-spotty" behaviour (involving more "uniform" distribution of turbulent fluctuations in time) for the large separation case, encompassing the entire spectrum of transition scenarios. The variation of the intermittency factor within the transition zone is seen to compare well with the Narasimha universal intermittency distribution. We find that although the time variation of velocity for large separation is non-spotty (or more "uniform"), the spanwise variation of velocity is spotty, showing a clear clustering of high-wavenumber fluctuations separated by quasi-laminar regions. Thus, all the cases exhibit spottiness in the transition zone with different manifestations. We present a physical cartoon for the transition scenarios for the attached and separated cases, using the ideas of vortex-wall interaction and instability of spanwise vortical structures. We find that concentrated breakdown is exhibited by all the three cases near the transition onset and the spot breakdown processes are broadly consistent with the postulates underlying the universal intermittency distribution.

---

\* Corresponding author: sdiwan@iisc.ac.in
[1] Current address: Department of Aerospace Engineering, Karunya Institute of Technology and Science, Coimbatore, Tamil Nadu, India – 641114. Email: ratnakumar@karunya.edu
[2] Current address: Associate Engineer, Caterpillar India Pvt Ltd, Bengaluru 560048. Email: karthikvnaicker@gmail.com

# 1. INTRODUCTION

Boundary layer transition has been a subject of research interest for several decades due to its fundamental importance and practical applications. The route to transition is not unique and is often influenced by leading-edge effects, pressure gradient and upstream disturbances [1-2]. In a boundary layer, the transition route can be categorised as "natural" and "bypass" [1-2]. The former is associated with the amplification of the Tollmien-Schlichting (T-S) waves, followed by their secondary instability and finally breakdown to turbulence, whereas the latter is found to "by-pass" the T-S route and is typically observed for high levels of background turbulence, e.g., when a boundary layer is subjected to free-stream turbulence (FST). For such a bypass scenario, low-frequency disturbances are admitted into the boundary layer through a process called "shear sheltering" that results in elongated spanwise modulations of the streamwise velocity called the streaks [3-4]. The streamwise streaks undergo rapid amplification as a result of a secondary instability and a subsequent breakdown marking the onset of transition [5-7].

There also exist conditions under which both natural and bypass transition scenarios can be present simultaneously. The relative importance of the two scenarios depends on the conditions favouring their development within the boundary layer. Furthermore, there can also be mutual interactions between T-S waves and streaks, and the resulting transition mechanism has been termed as the "mixed mode" of transition [8-12]. Bose and Durbin [12] studied such a scenario in zero pressure gradient (ZPG) boundary layers. In their simulations, both a TS wave and FST were injected at the inflow, and transition occurred through their interaction. The mixed-mode mechanism is more commonly observed for FST induced transition in the presence of an adverse pressure gradient (APG), since an APG promotes growth of instability waves due to the presence of an inflection point in the mean velocity profiles. Some early studies to investigate the transition process in attached APG boundary layers were due to Walker and Gostelow [13] and Gostelow *et al*. [14]. In a recent numerical simulation of Bose *et al*. [11], an attached APG boundary layer was subjected to varying levels of FST and the mixed-mode of transition was investigated. At low levels of FST (0.1%), Bose *et al*. [11] found existence of both T-S waves and streaks, and the onset of transition was observed through a secondary instability of the T-S wave. For higher levels of FST (1 - 2%), the bypass transition was found to be dominant although instability waves were still apparent, with their amplitudes comparable to the streak amplitudes. Another example of the mixed-mode transition was reported by Zaki *et al*. [15], on the pressure side of a compressor blade under low levels of FST, wherein they obtained an attached boundary layer throughout the blade chord.

An important consequence of the presence of a sufficiently strong APG is the boundary layer separation, which can have a significant influence on the underlying transition mechanism. There have been several studies [16-25] investigating the transition process in separated shear layers, in particular pertaining to the natural modes of transition (FST ≤ 0.1%). The inviscid instability associated with the inflectional base velocity profiles, was seen to be the primary instability mechanism for laminar separation bubbles (LSBs) [16-25]. The movement of inflection point away from the wall due to boundary layer separation was found to result in significantly high growth rates as compared to an attached APG boundary layer. The instability of a separated shear layer is usually termed as "Kelvin-Helmholtz (KH)" instability (which is typical of a free shear layer) [16-21]. However, Diwan and Ramesh [23] have shown that the effect of wall on the instability dynamics cannot be neglected and the origin of the inflectional instability can be traced back to the upstream attached boundary layer. The subsequent stages of transition in a separation bubble undergoing natural transition include formation of spanwise vortical rollers and their breakdown into smaller-scales through a secondary instability, which eventually leads to turbulence close to or downstream of the reattachment point of the separation bubble [20-32].

There have also been studies to understand the effect of elevated levels of FST (FST > 0.1%) on the transition mechanism in an LSB, wherein both inflectional and bypass mechanisms can be expected to be present [24-28]. Mutual interactions between the boundary layer streaks and spanwise shear-layer rollers can result in significant changes in the dynamics of transition in the separated shear layer [25, 27, 30]. The experiments of Haggmark [28] on the flow developing on a flat plate under the influence of an APG at elevated FST levels showed the existence of large amplitude low frequency streaks in the attached boundary layer as well as in the separated shear layer. Recent experiments of Dellacasagrande *et al.* [31] and Istvan and Yarusevych [25] on transition in separated shear layers for varying Reynolds numbers and FST indicated that the inflectional instability was present even at elevated FST. On the other hand, the numerical simulations of McAuliffe and Yaras [29] at elevated FST (1.45% at the onset of separation), illustrated that the instability mechanism leading to shear-layer rollup was bypassed as a result of streamwise streaks strongly interacting with the separated shear layer. The direct numerical simulations of Blazer and Fasel [26] and Hosseinverdi and Fasel [27] investigated in detail the effect of varying levels of FST on transition in a separated shear layer. They found that the dominant transition mechanism can be either inflectional instability or the transient growth associated with streamwise streaks. They also presented cases wherein the interaction between the two mechanisms resulted in a breakup of the spanwise rollers and a development of chaotic 3D structures [27].

The onset of transition in an attached or separated boundary layer marks the beginning of the "transition zone", which is a finite region over which skin friction increases gradually from a laminar value to a turbulent value. An important quantity that characterises the transition zone is the "intermittency factor" ($\gamma$), originally proposed by Emmons [38], which is defined as the fraction of the time the flow is turbulent. The intermittency factor varies from 0 to 1, where $\gamma = 0$ represents the onset of transition and $\gamma = 1$ represents the end of transition i.e., the onset of turbulence. Narasimha [39] proposed a distribution for $\gamma$ within the transition zone in a ZPG boundary layer. This distribution was successfully applied to transitional flows subjected to surface roughness, FST [40] and favourable pressure gradient [41] and was therefore termed a "universal" intermittency distribution. Walker and Gostelow [13] measured the transition zone in an attached APG boundary layer and found that their $\gamma$–distribution matched favourably with this universal distribution. Note that in all the cases of the attached boundary layer mentioned above, the transition zone is typically characterised by the appearance of "turbulent spots", which are the "islands" of turbulent flow surrounded by quasi-laminar regions [38]. Moreover, the universal $\gamma$–distribution of Narasimha [39] is based on the "concentrated breakdown hypothesis", which states that the turbulent spots are generated at a preferred streamwise location randomly in time and spanwise direction; another key postulate for obtaining this distribution is the Poisson process for the spot generation rate [41].

There have been attempts to check if the universal $\gamma$–distribution works for the transition zone in a separated shear layer [19, 42-43]. A relevant question in this connection is whether turbulent spots are observed in the separated flow transition. McAullife and Yaras [44] found that for a short separation bubble, the velocity signals in the transition zone exhibited turbulent spots (we also show this for the small bubble in our numerical simulation). However, majority of the studies, especially on the moderate to large sized bubbles, do not find appearance of distinct spots in the transition zone [16-28]. Instead, the velocity time traces indicate an increase of high frequency fluctuations more or less "uniformly" over the entire time duration (without any obvious clustering of high frequencies in the form of spots). Notwithstanding this fact, the intermittency distribution in the transition zone of the moderate to large separation bubbles is found to compare reasonably well with Narasimha's universal distribution [19, 42-43], which is an intriguing result. This point has been recognised and discussed in the past [19, 45] but has not been investigated in sufficient detail to our best knowledge.

In this work, we attempt to address this question by carrying out a direct numerical simulation (DNS) of an incompressible flow over a flat plate subjected to an APG. Three inlet Reynolds numbers ($Re_{\delta_{in}^*} = U_{ref}\,\delta_{in}^*/\nu$) are chosen such that at the highest $Re_{\delta_{in}^*}$ (158.7) the boundary layer is attached, at an intermediate $Re_{\delta_{in}^*}$ (105.8) a small separation bubble is obtained and at low $Re_{\delta_{in}^*}$ (79.2) a large bubble is obtained. Here $U_{ref}$ is the incoming free-stream velocity, $\delta_{in}^*$ is the displacement thickness at the inlet and $\nu$ is the kinematic viscosity. The simulation of attached and two separated flow cases in the same computational setting allows us to investigate the differences in the transition processes in the attached and separated boundary layers, in particular in relation to the question of turbulent spots. To the best of our knowledge, such a simulation has not been reported in the past. There have been studies which have used elevated levels of FST to eliminate LSB and get an attached boundary layer [25 - 28]. However, the present simulation setting (involving fixed FST and changing *Re*) and objectives are quite different from these studies. In particular, we choose a moderately low value of FST intensity of 0.3% (at separation location) for our simulations, for reasons listed below:

- There are several studies that describe the effect of moderate to high FST (> 0.5%) on the flow field and transition mechanism in separation bubbles [17-20, 24-29]. However, studies for 0.1% < FST < 0.5% have been relatively few [25, 29]. It was therefore, thought worthwhile focussing on the low range of FST and bring out the interaction between the instability waves and streamwise streaks as the boundary layer changes its character from attached to separated.
- The FST in standard wind tunnels ranges from 0.1 – 0.4% (except for the specially designed wind tunnels where it is < 0.1%). The results from the present simulations could therefore be used to compare with the experiments on separated flow transition conducted in such wind tunnels. For example, the transition study of Walker and Gostelow [13], Istvan and Yarusevych [25] and Hatman and Wang [65-66] report measurements in a wind tunnel with turbulence intensity of 0.3% - 0.45%.
- The turbulence level of 0.3% is also relevant for practical situations, an example being an autonomous underwater glider moving in deep sea, which does not experience large background turbulence that is typically present near the sea surface/sea shore [46]. An un-manned air vehicle (UAV) flying at moderately high altitudes can also be expected to experience a relatively weak level of atmospheric turbulence.

In the following, we present a detailed analysis of the boundary layer structures in pre-transitional and transitional regions for the three simulated cases, and also perform a time-series analysis (using Fourier and wavelet transforms) on the velocity signals, to relate the features in the latter to those in the former. We find that, for attached as well as separated cases, the pre-transitional region is characterised by a "mixed-mode" instability, involving contributions from instability waves and streamwise streaks, with a breakdown of spanwise rollers marking the onset of transition in each case. Within the transition zone, the intermittency variation for all the three cases matches well with the universal $\gamma$–distribution of Narasimha [39] (consistent with the literature), even though the velocity time traces for large separation bubble do not show distinct turbulent spots. We also analyse the spanwise variation of streamwise fluctuating velocity and find that there is clustering of high-wavenumber fluctuations (akin to turbulent spots) in these signals even for the large separation case. By interpreting the "uniformly" appearing turbulent fluctuations as tailgating among turbulent spots, we discuss plausible reasons why the intermittency distribution for large separation matches reasonably well with the Narasimha universal distribution.

The paper is organized as follows: Section II describes the numerical method used and provides computational details. The characterisation of the pre-transitional region in terms of instability waves and streamwise streaks is presented in Section III. Section IV presents the intermittency calculation and time-frequency analysis of the velocity signals, in attached and separated cases. A physical cartoon of the spot breakdown scenarios is proposed in Section V and conclusion is presented in Section VI.

## II. NUMERICAL METHOD AND COMPUTATIONAL DETAILS

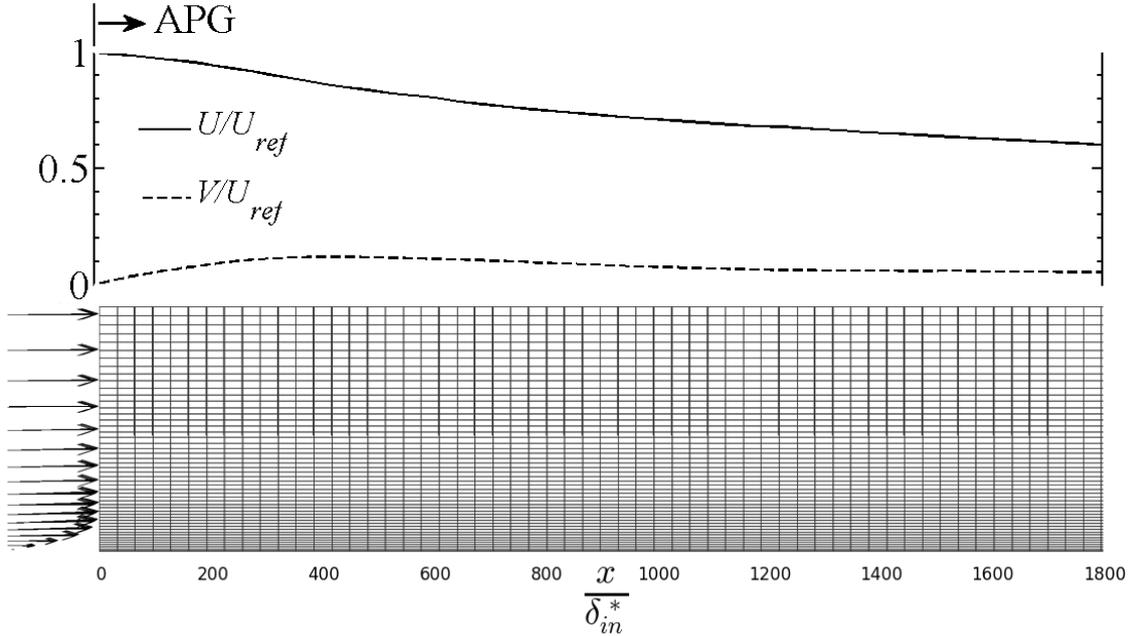

FIG. 1. Computational domain showing the inlet velocity profile and the top boundary condition (*U* and *V* distribution). The computational grid is also shown. *U*: Mean streamwise velocity; and *V*: Mean wall-normal velocity.

In the present work, the three-dimensional, unsteady and incompressible Navier-Stokes equations are solved numerically. The equations are non-dimensionalised using incoming free-stream velocity as the reference velocity $(U_{ref})$, and the displacement thickness at inlet $(\delta_{in}^*)$ as the reference length. The equations are solved using the fractional step method of Kim and Moin [47], which is based on the projection method proposed by Chorin [48]. All spatial derivatives are discretised by using the second order central difference scheme. Integration in time is explicit for all terms except for the viscous term in the wall-normal direction, for which the semi-implicit Crank-Nicholson scheme is used. The Courant–Friedrichs–Lewy (CFL) number is checked every step to ensure the time-step size is appropriate to maintain CFL less than one. The time integration for the first sub-step is carried out using the third-order Runge-Kutta method [49]. For the streamwise (*x*) and spanwise (*z*) directions, which have uniform grid spacing, the eigenfunction expansion method of Swarztrauber [50] is used. The wall-normal direction (*y*) has non-uniform grid spacing with an expansion ratio of 1.03 to ensure good near-wall resolution of the large velocity gradients due to shear. In this case, a tri-diagonal matrix is obtained which is inverted using the Thomas algorithm. For the present simulations, we have used a variant of the solver [51] developed by Patwardhan [52]. For the validation of the code the reader is referred to [52].

The governing equations are solved in a cuboidal domain in Cartesian coordinates (Fig. 1). A staggered grid approach is used, with velocity components defined on the cell faces and the pressure at the cell centres [52]. Appropriate boundary conditions are prescribed on the faces of the cuboid. The bottom face represents a flat plate where no-slip and no-penetration conditions are prescribed. Since the time-mean flow is two dimensional and the mean spanwise velocity is zero, periodic boundary conditions are prescribed on the lateral (side) walls. The condition at the exit is not known apriori and evolves with the upstream flow. For the turbulent flow which develops after reattachment for the separation bubble cases, the suitable boundary condition is found to be the convective outflow boundary condition, as it allows vortical structures to move out of the domain with minimal distortion [53]. The same boundary condition is also found sufficient for the attached case. At the inlet face, velocity components are prescribed as Dirichlet conditions. For the streamwise velocity component, a Blasius profile is provided which is superimposed with random perturbations (Fig. 1). The level of inlet perturbations is such as to ensure FST of 0.3% for all the cases specified, at the minimum skin-friction location when the flow is attached and at the onset of separation for separated flow cases. The Dirichlet boundary condition specified on the top wall for the velocity components controls the pressure gradient imposed on the flow (Fig. 1). The streamwise ($U$) and wall-normal ($V$) velocity distributions on the top wall were obtained by carrying a preliminary analysis in ANSYS Fluent [51]. The exercise consisted of simulating the streamwise pressure gradient used in the experiments of [23] using the SST $k$-$\omega$ turbulence model. The $U$ & $V$ components at the required height from the wall are extracted from the ANSYS simulation and prescribed as the top-wall boundary condition for the present DNS. The same $U$ & $V$ distributions are used for all the three cases simulated herein. Note that this method of prescribing the top-wall boundary condition (i.e., Dirichlet condition for $U$ & $V$) has been used before in the literature [32].

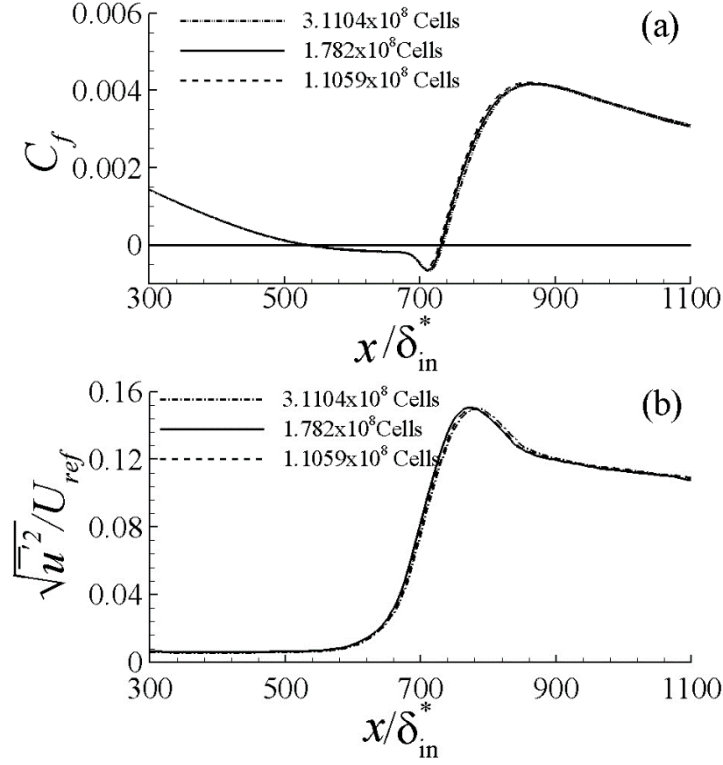

FIG. 2. Grid independence exercise illustrating the streamwise variation of (a) skin friction and (b) r.m.s. of streamwise velocity fluctuations.

The computational grid used in this study is generated following grid convergence studies and a comparison of the grid resolution with other works. A grid of size 2400 x 300 x 240 ($L_x = 1800\delta_{in}^*$; $L_y = 350\delta_{in}^*$ and $L_z = 180\delta_{in}^*$) is selected for simulating all the cases considered here with total number of grid points equal to $1.782\times10^8$; here $L$ is the length of the domain. Figure 2 shows the grid independence study comparing the skin-friction coefficient $\left(C_f = \tau_w/0.5\rho U_{ref}^2\right)$, and the r.m.s. of streamwise fluctuating velocity $\left(\sqrt{\overline{u'^2}}\right)$ for $Re_{\delta_{in}^*} = 79.2$. Here, $\tau_w$ is wall shear stress, $\rho$ is density, the dashed quantities indicate fluctuations about the mean and the overbar indicates time averaging. As can be seen from Fig. 2, $C_f$ and $\frac{\sqrt{\overline{u'^2}}}{U_{ref}^2}$ show a negligible variation for change in the grid size by a factor 3. It was therefore thought adequate to use the grid with $1.782\times10^8$ cells. Note that the streamwise variations of $C_f$ and $\sqrt{\overline{u'^2}}$ in Fig. 2 are consistent with the standard results reported by the computational studies on separation bubbles [26-27, 32]. Next, we determine the grid resolution in "wall units" and compare it with the previous studies (Table 1). The wall units are defined as $\Delta y^+ = \Delta y u_\tau/\nu$, where $\Delta y$ is the grid spacing and $u_\tau = \sqrt{\tau_w/\rho}$ is the friction velocity. The values included in Table 1 are evaluated at the maximum skin-friction location near reattachment region, which is the most difficult region of the flow to resolve [32]. The grid resolution in the present simulations is finer than most of the studies included in Table 1 and is comparable to the recent work of Hosseinverdi and Fasel [27]. To obtain the statistics, the time-series data has been averaged over a non-dimensional time (= $tU_{ref}/\delta_{in}^*$) of 15000. This averaging time showed a good statistical convergence and averaging for a longer duration did not result in further improvement. The time series data shown in Section IV are for a somewhat shorter duration than this averaging interval ($tU_{ref}/\delta_{in}^*$ = 2500-6000).

TABLE 1: Comparison of grid resolution in wall units with that reported in the literature ("N" is the number of grid points for $y^+<9$.)

| Case | $\Delta x^+$ | $\Delta y^+$ at $y^+ = 9$ | $\Delta z^+$ | $N(y^+<9)$ |
|---|---|---|---|---|
| Alam and Sandham [32], case 3DF-B | 14.26 | 0.87 | 6.3 | 17 |
| Jones et al. [36], case 3DF | 3.36 | $\geq 1$ | 6.49 | $\leq 9$ |
| Marxen and Henningson [37], case reso1 | 6.53 | 0.94 | 11.06 | 10 |
| Balzer and Fasel [26] | 5.6 | 0.9 | 6.15 | 18 |
| **Present Work** | **2.87** | **0.6** | **3.07** | **20** |
| Hosseinverdi and Fasel [27] (with FST) | 1.58 | 0.44 | 2.71 | 25 |

## III. FLOW CHARACTERIZATION AND INSTABILITY MECHANISMS

### 1. Mean and Perturbation Flow Fields

Figure 3 shows streamwise variation of the skin-friction coefficient ($C_f$) (Fig. 3(a)) and coefficient of wall pressure $\left(C_p = \frac{p-p_{ref}}{0.5\rho U_{ref}^2}\right)$ (Fig. 3(b)) for the attached flow case ($Re_{\delta_{in}^*}$ = 158.7) and the two separated flow cases ($Re_{\delta_{in}^*}$ = 105.8 and 79.2). At $Re_{\delta_{in}^*}$ = 158.7, the flow remains attached despite the adverse pressure gradient; here, $C_p$ shows a region of sharp decrease for $x/\delta_{in}^* > 500$ and $C_f > 0$ for all $x/\delta_{in}^*$. As $Re_{\delta_{in}^*}$ decreases, $C_f$ shows zero crossings (Fig. 3(a)) indicating the streamwise extent of the separated flow region. The first zero crossing point of $C_f$ indicates the onset of separation, whereas the second crossing point indicates the reattachment location [32]. Furthermore, a flat skin-friction distribution followed by a negative peak in $C_f$ for $Re_{\delta_{in}^*} = 105.8$ and $79.2$ describe the "dead-air" and reverse-flow vortex regions respectively; see Fig. 3(a). The $C_p$ distributions for these cases show a weaker variation in the initial portion of the separation bubble followed by a sharp recovery close to the reattachment point (Fig. 3(b)). These are characteristic features of a laminar separation bubble, which are captured well in the present simulations [21, 26, 32]. Following the literature [12, 29-30, 32], the onset and end of transition are defined as locations where $C_f$ is minimum and maximum respectively. The streamwise locations indicating the separation, transition and reattachment points are listed in Table 2; also included are the maximum height of the separation bubbles $\left(\text{at } Re_{\delta_{in}^*} = 105.8 \text{ and } 79.2\right)$ and the streamwise locations of the maximum height. It is interesting to note that the onset of transition coincides with the maximum-height location for both the separated flow cases.

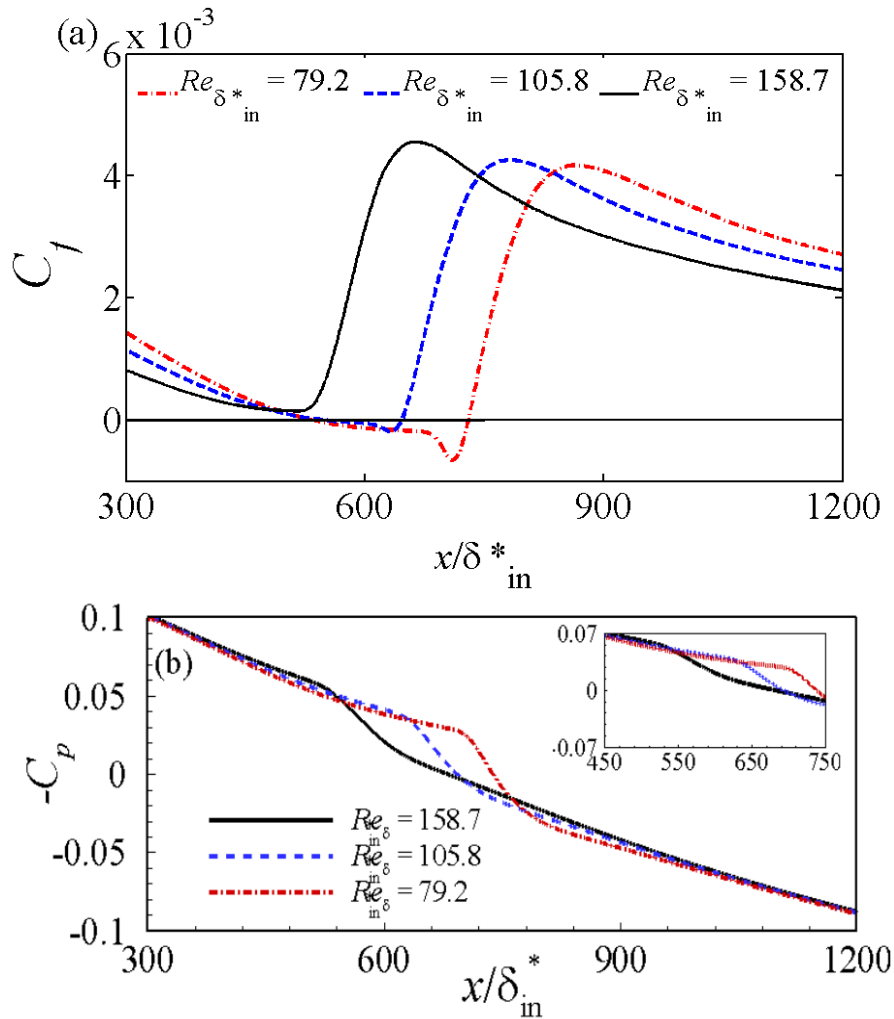

FIG. 3. (a) Mean skin-friction coefficient ($C_f$) and (b) mean coefficient of wall pressure ($C_p$) plotted against normalised streamwise distance for the three simulated cases.

TABLE 2: Streamwise locations corresponding to different features in the flow for the three simulated cases. Here $x_s$ – separation point; $x_{to}$ – onset of transition; $x_r$ – reattachment point; $x_{te}$ – end of transition; $x_M$ – streamwise location of maximum height of the bubble and $h$ - maximum height of the bubble. All distances scaled on $\delta_{in}^*$. $Re_{\delta_s^*} = U_s \delta_s^*/\nu$; $U_s$ and $\delta_s^*$ respectively are the local freestream velocity and displacement thickness at $x_s$.

| $Re_{\delta_{in}^*}$ | $(x_s)$ | $(x_{to})$ | $(x_r)$ | $(x_{te})$ | $(x_M)$ | $(h)$ | $Re_{\delta_s^*}$ |
|---|---|---|---|---|---|---|---|
| 158.7 (attached) | - | 524 | - | 660 | - | - | - |
| 105.8 | 550 | 620 | 650 | 774 | 620 | 1.0 | 524.0 |
| 79.2 | 530 | 690 | 730 | 857 | 690 | 3.0 | 429.5 |

The changes in the flow field with $Re_{\delta_{in}^*}$ in the central plane ($z/\delta_{in}^* = 90$) are illustrated in Fig. 4, where the velocity vectors are superimposed on the contours of the Reynolds shear stress $\left(-\overline{u'v'}/U_{ref}^2\right)$; here $v$ is the wall-normal velocity component. The $-\overline{u'v'}/U_{ref}^2$ contours less than 0.001 are not shown for better representation [35]. At $Re_{\delta_{in}^*} = 158.7$, the boundary layer is attached and the maximum values of the Reynolds stress are closer to the wall. The mean separation bubbles obtained at $Re_{\delta_{in}^*} = 105.8$ and 79.2 are indicated by dividing streamlines (Fig. 4(b) and (c)). The separation locations for these two cases are not very different, whereas the maximum height and reattachment locations show a significant increase as $Re_{\delta_{in}^*}$ decreases from 105.8 to 79.2 (Fig. 4, Table 2). The transition point moves downstream with decreasing $Re_{\delta_{in}^*}$ as expected (Table 2) and for the separated cases this results in a longer bubble for $Re_{\delta_{in}^*} = 79.2$ as compared to $Re_{\delta_{in}^*} = 105.8$. For these two Reynolds numbers, the Reynolds shear stress is seen to peak near the reattachment location, consistent with the literature [18, 20]. Note that both the separation bubbles shown in Fig.4, are classified as "short" bubbles as per the Diwan-Chetan-Ramesh criterion [33]. In the separation bubble literature, the bubbles are classified as "short" and "long" based on the departure of the actual wall pressure distribution from the imposed inviscid pressure distribution [33, 34]; a short bubble exhibits only a slight deviation from the inviscid distribution. According to the Diwan-Chetan-Ramesh criterion a bubble is classified as short if parameter $P = \frac{h^2}{\nu} \frac{\Delta U}{\Delta x} > -28$. Here $\Delta x = x_r - x_s$ and $\Delta U$ is the free-stream velocity difference between the separation and reattachment points. For the present cases, we find $P = -2$ for $Re_{\delta_{in}^*} = 105.8$ and $P = -12$ for $Re_{\delta_{in}^*} = 79.2$.

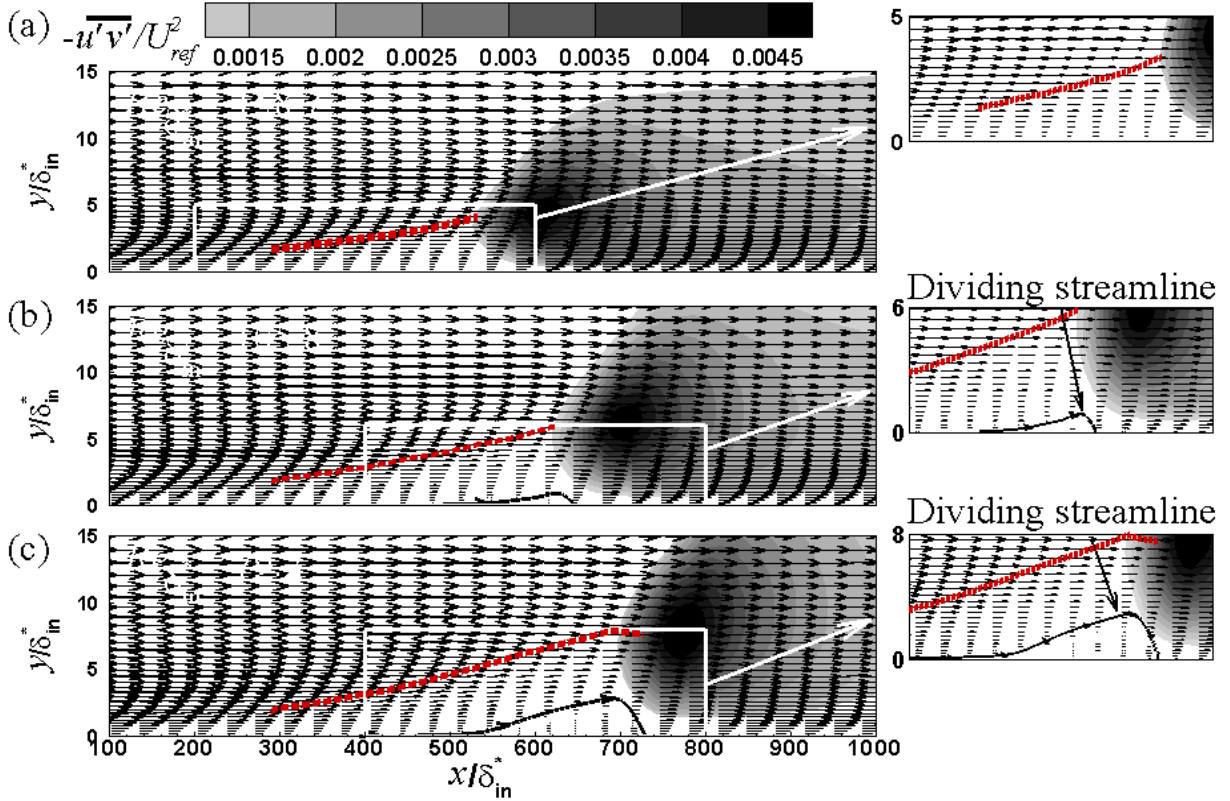

FIG. 4. Changes in the flow field with decreasing $Re_{\delta_{in}^*}$: (a) $Re_{\delta_{in}^*} = 158.7$; (b) $Re_{\delta_{in}^*} = 105.8$ and (c) $Re_{\delta_{in}^*} = 79.2$. Normalised time-averaged velocity vectors in the centre-plane are superimposed on the contours of the Reynolds shear stress $\left(-\overline{u'v'}/U_{ref}^2\right)$. Dashed line indicates the locus of inflection points until the onset of transition, while the solid line is the dividing streamline for the bubbles.

The streamwise distribution of the maximum in $\overline{u'^2}$ is shown in Fig. 5. For all the three cases, $\overline{u'^2}$ exhibits a mild growth in the initial region ($x/\delta_{in}^* < 450$ for the attached case and $x < x_s$ for the separated cases), followed by a strong growth until the onset of transition ($x_{to}$). The region immediately preceding $x_{to}$, shows a near exponential growth, seen as an approximate straight-line behaviour of $\overline{u'^2}$ in the semi-log plot in Fig. 5. The presence of exponential growth implies linear modal instability, which is expected due to the presence of an inflection point in the mean velocity profiles [23]. On the other hand, the weak disturbance growth in the initial region seen in Fig. 5 can be attributed to the transient growth of the streamwise streaks as will be discussed in the next section. Downstream of the transition location, the $\overline{u'^2}$ curves deviate from the exponential trend and reach saturation amplitudes further downstream (Fig. 5).

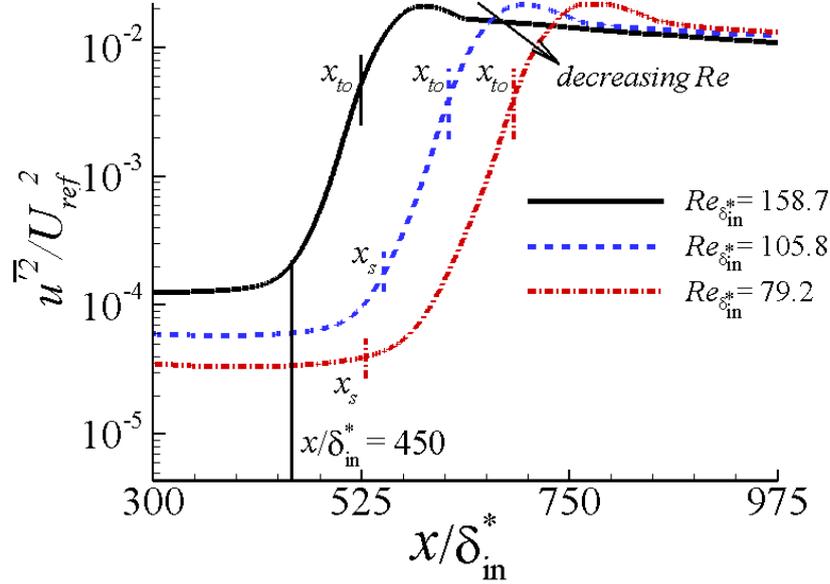

FIG. 5. Streamwise evolution of the maximum in $\overline{u'^2}$ for the three simulated cases.

#### 2. Instability Mechanisms and Onset of Transition

The wall-normal profiles of the streamwise turbulence intensity are plotted in Fig. 6 for the three cases. Figure 6(a) shows that at $x/\delta^*_{in} = 450$, (a representative location where the boundary layer is attached for all three cases), the $\overline{u'^2}$ profile peaks close to $y/\delta^* = 1.3$ for $Re_{\delta^*_{in}} = 79.2$ and 105.8, whereas it peaks close to $y/\delta^* = 1$ for the attached case $\left(Re_{\delta^*_{in}} = 158.7\right)$. Here $\delta^*$ is the local displacement thickness, i.e., $\delta^* = \delta^*(x)$. It is well known that, when a zero-pressure-gradient boundary layer is subjected to FST, the disturbance amplitude peaks at $y/\delta^* = 1.3$ in the pre-transitional region [54]. Balzer and Fasel [26], in their simulations with FST = 0.5%, found that the peak in disturbance amplitude was close to $y/\delta^* = 1.3$ even under an adverse pressure gradient (in the attached boundary layer upstream of separation). Furthermore, they found that the typical shape of the disturbance amplitude within the boundary layer was similar to that predicted by Luchini [54] using the optimal perturbation theory. The $\overline{u'^2}$ distribution in the present simulations for $Re_{\delta^*_{in}} = 105.8$ and 79.2 has the same qualitative shape as reported in Balzer and Fasel [26]. This suggests that the dominant mechanism present at $x/\delta_{in}^* = 450$, at these Reynolds numbers, is the lift-up effect of the streamwise streaks [54]. This inference is supported by the visualization of $u'$ contours in Fig. 7 (c) and (e), which clearly shows alternative bands of low-and high-speed streaks that extend beyond $x/\delta_{in}^* = 450$ for $Re_{\delta_{in}^*}$ = 105.8 & 79.2. The contour plots in Fig. 7 are presented for the *x-z* plane (parallel to the wall) corresponding to $y/\delta_{to} \approx 0.5$, where $\delta_{to}$ is the 99% boundary layer thickness at the transition, $x_{to}$. This choice of *y* location was motivated by the work of Bose *et al.* [11]. (Fig. 8 shows the *x-z* contour plots at another wall normal location, $y/\delta_{to} \approx 1$.) Note that the streamwise streaks are also present at $Re_{\delta_{in}^*}$ = 158.7 (Fig. 7(a)), but at $x/\delta_{in}^* = 450$, it is modulated by spanwise bands of streamwise velocity; these are seen more clearly in *v'* distribution (Fig. 7 (b)). The spanwise bands in *u'* and *v'* components can be associated with the 2D instability waves, which were also reported by Bose *et al.* [11] in their simulation of an attached APG boundary layer. These instability waves typically correspond to the modal solutions of the Orr-Sommerfeld (O-S) equation cast as an eigenvalue problem [55].

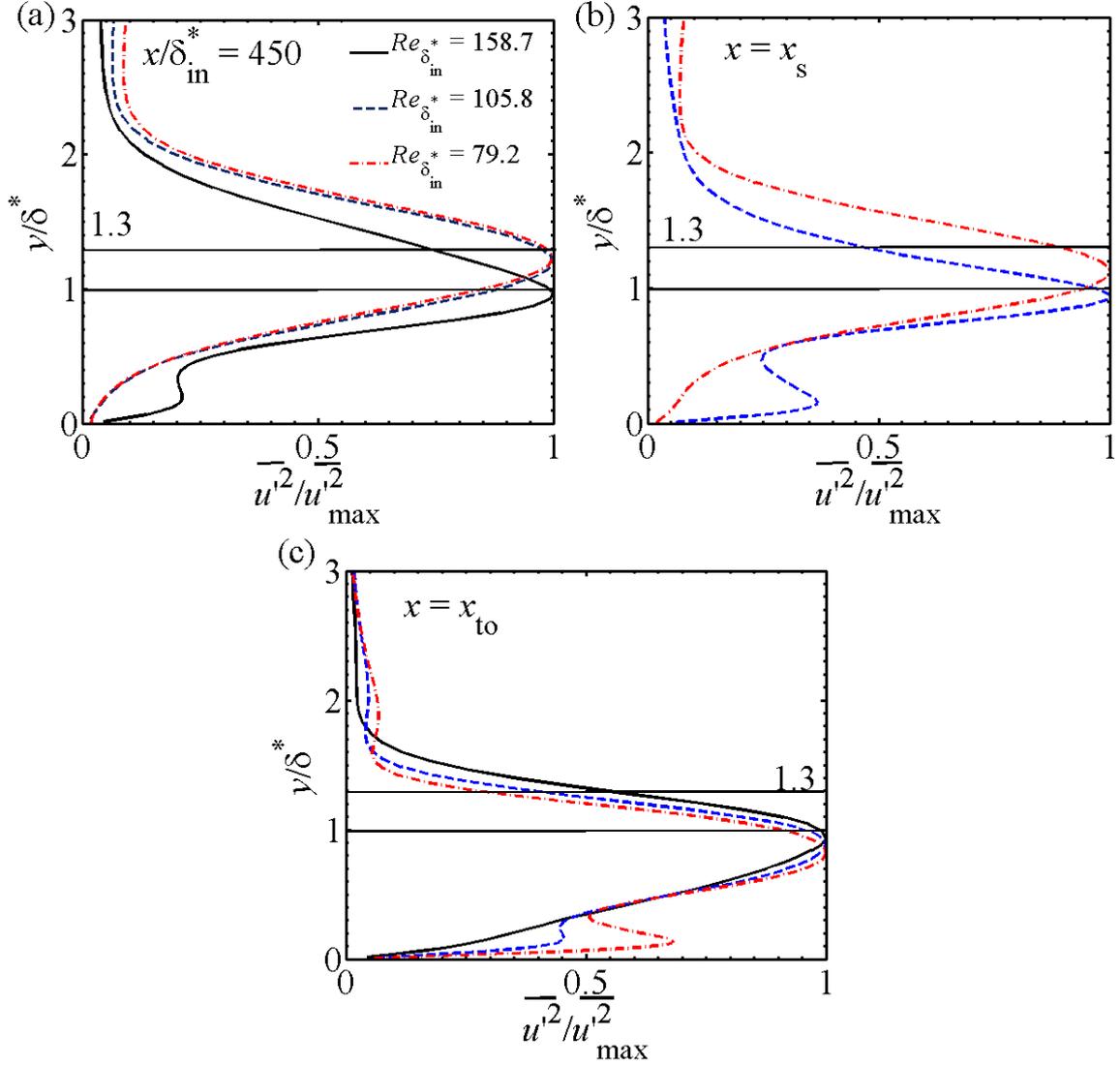

FIG. 6. Wall normal variation of $\overline{u'^2}/\overline{u'^2}_{max}$ at select streamwise locations for varying $Re_{\delta^*_{in}}$: (a) $x/\delta^*_{in} = 450$; (b) at the onset of separation $(x_s)$ and (c) at the onset of transition $(x_{to})$.

Referring back to Fig. 6(a), we see that the shape of $\overline{u'^2}$ distribution is qualitatively different at $Re_{\delta^*_{in}} = 158.7$ (attached flow), as compared to that at two lower $Re_{\delta^*_{in}}$. As mentioned earlier, at $Re_{\delta^*_{in}} = 158.7$, $\overline{u'^2}$ peaks at $y/\delta^* = 1$ (instead of $y/\delta^* = 1.3$), which is close to the location of the inflection point in the pre-transitional region. (For all the three cases, the locus of $\delta^*$ is found to be close to the locus of inflection point in the pre-transitional region; see also [26]). Furthermore, there also exists a local peak in $\overline{u'^2}$ closer to the wall (Fig. 6(a)) at $Re_{\delta_{in}^*}$ = 158.7. A peak in $\overline{u'^2}$ close to the inflection point and a "double hump" shape of the $\overline{u'^2}$ profile are typical features of the inflectional instability mechanism [23]. These observations support our earlier inference that for $Re_{\delta_{in}^*}$ = 105.8 and 79.2, the dominant mechanism at $x/\delta_{in}^*$ = 450 is the lift-up effect of streamwise streaks (although weak spanwise bands corresponding to instability waves are apparent (Fig. 7(d) & (f)). On the other hand, for the highest $Re_{\delta_{in}^*}$, the inflectional instability mechanism has already become dominant at this streamwise location. This is consistent with the evolution of $\overline{u'^2}$ shown in Fig. 5; at $Re_{\delta_{in}^*}$ = 105.8 and 79.2, the growth in the disturbance energy is weak at $x/\delta_{in}^*$ = 450, whereas at $Re_{\delta_{in}^*}$ = 158.7, a much higher growth rate is seen at this location typical of inflectional instability. The effect of inflectional instability, for $Re_{\delta_{in}^*}$ = 158.7, is clearly manifested in the downstream of $x/\delta_{in}^*$ = 450, wherein a strong exponential growth is seen; see Fig. 5. This observation is consistent with the observations of Bose *et al.* [11] on an attached APG boundary layer for FST of 0.1%. They also found presence of streaks and instability waves within the boundary layer, with $\overline{u'^2}$ showing an exponential variation in the streamwise direction.

For the two lower $Re_{\delta_{in}^*}$, the $\overline{u'^2}$ profile shapes at the separation location show deviation from those at $x/\delta_{in}^*$ = 450 (Fig. 6 (a) & (b)). For $Re_{\delta_{in}^*}$ = 105.8, the mode shape looks typical of inflectional instability with a peak close to $y/\delta^*$ = 1 (or point of inflection) and a secondary peak closer to wall. This is again reflected in the $v'$ contours (Fig. 7(d)), wherein the 2D instability waves are clearly seen at the separation point at $Re_{\delta_{in}^*}$ = 105.8. For $Re_{\delta_{in}^*}$ = 79.2, the mode shape (Fig. 6 (b)) is a combination of those typical of lift-up mechanism and inflectional instability, as streamwise streaks are seen extended beyond the separation point at this $Re_{\delta_{in}^*}$ with instability waves becoming dominant further downstream (Fig. 7(e) and (f)). The persistence of streaks beyond separation and relatively late onset of instability at $Re_{\delta_{in}^*}$ = 79.2 compared to $Re_{\delta_{in}^*}$ = 105.8 could be attributed to the Reynolds number effect.

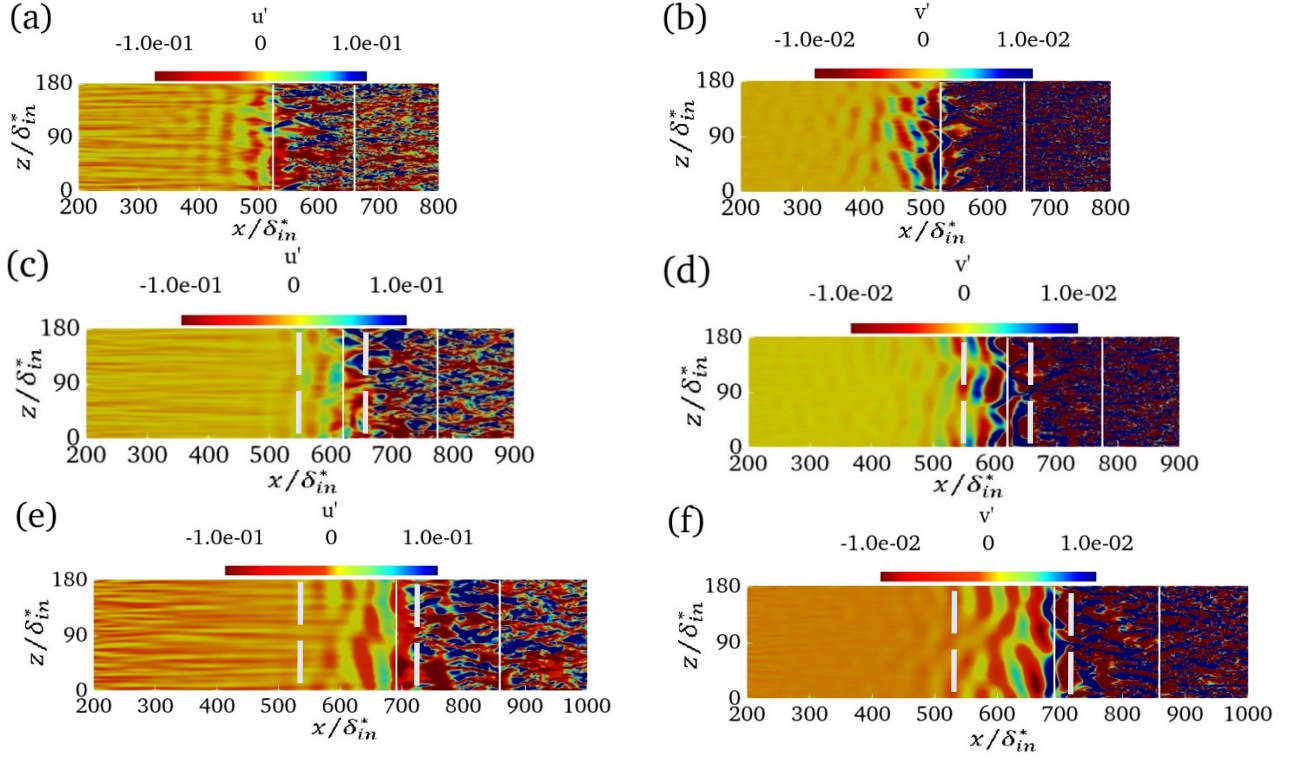

FIG. 7. Instantaneous contours of $u'$ and $v'$ in $x$-$z$ plane at $y/\delta_{to} = 0.5$. Top row: $Re_{\delta_{in}^*} = 158.7$; middle row: $Re_{\delta_{in}^*} = 105.8$ and bottom row: $Re_{\delta_{in}^*} = 79.2$ (left pane contours of $u'$ and right pane contours of $v'$). Solid white lines indicate onset and end of transition, whereas dashed white lines indicate separation and reattachment points.

At the location of transition onset ($x_{to}$), the mode shapes for the two separated cases are typical of inflectional instability (Fig. 6 (c)), whereas that for the attached case is somewhat different. While, for the attached case, the $\overline{u'^2}$ distribution peaks close to $y/\delta^* = 1$, a secondary peak closer to the wall is absent. This suggests that the lift-up mechanism continues to be active for $Re_{\delta_{in}^*} = 158.7$ until the onset of transition and interacts actively with the inflectional instability mechanism. Therefore, the $\overline{u'^2}$ profile at $x_{to}$ for the attached case shares the features of both the mechanisms as seen in Fig. 6(c) (similar to what is seen for $Re_{\delta_{in}^*} = 79.2$ at $x_s$; Fig. 6(b)). The evidence for this is seen in Fig. 7, which shows that the spanwise coherence of disturbances (typical of inflectional instability) at $x_{to}$ is much stronger for the two separated cases as compared to the attached case. For the attached case (Fig 7 (a) & (b)) oblique structures are visible at the onset of transition which suggest an interaction between streamwise streaks and instability waves.

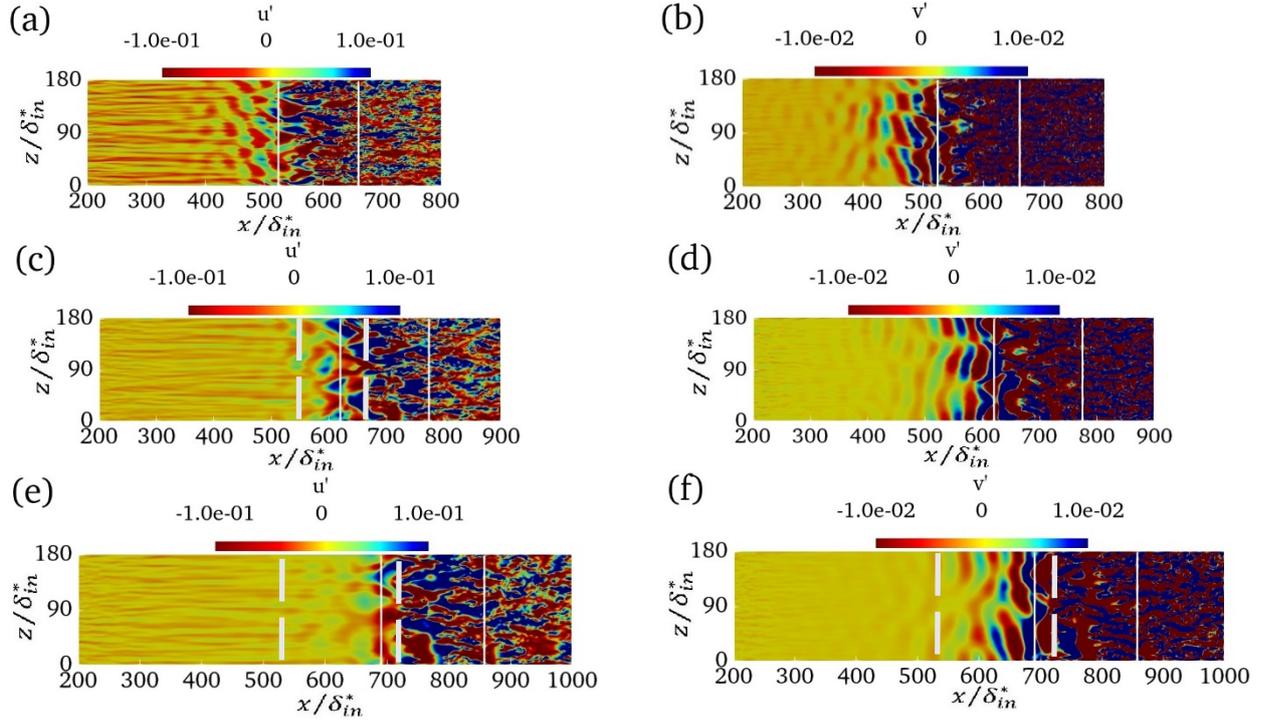

FIG. 8. Instantaneous contours of $u'$ and $v'$ in *x-z* plane at $y/\delta_{to} \approx 1$. Top row: $Re_{\delta_{in}^*}$ = 158.7, middle row: $Re_{\delta_{in}^*}$ = 105.8 and bottom row: $Re_{\delta_{in}^*}$ = 79.2 (left pane contours of $u'$ and right pane contours of $v'$). Solid white lines indicate onset and end of transition, whereas dashed white lines indicate the separation and reattachment points.

Figure 8 shows the $u'$ and $v'$ contour plots in the *x-z* plane at $y/\delta_{to} \approx 1$, at the location of transition onset. A similar pattern of streamwise streaks and spanwise instability bands (and their interaction) is seen at $y/\delta_{to} \approx 1$ (Fig. 8) to that seen at $y/\delta_{to} \approx 0.5$ (Fig. 7). This implies that the features we have described here are typical of the entire boundary layer, for each of the three cases. The foregoing discussion suggests that the primary instability mechanism for the attached and separated cases could be termed the "mixed-mode" instability, wherein both the transient growth associated with the lift-up effect and the modal inflectional instability are present. The relative contribution of the two instability modes varies for the three cases, as the transition onset is approached.

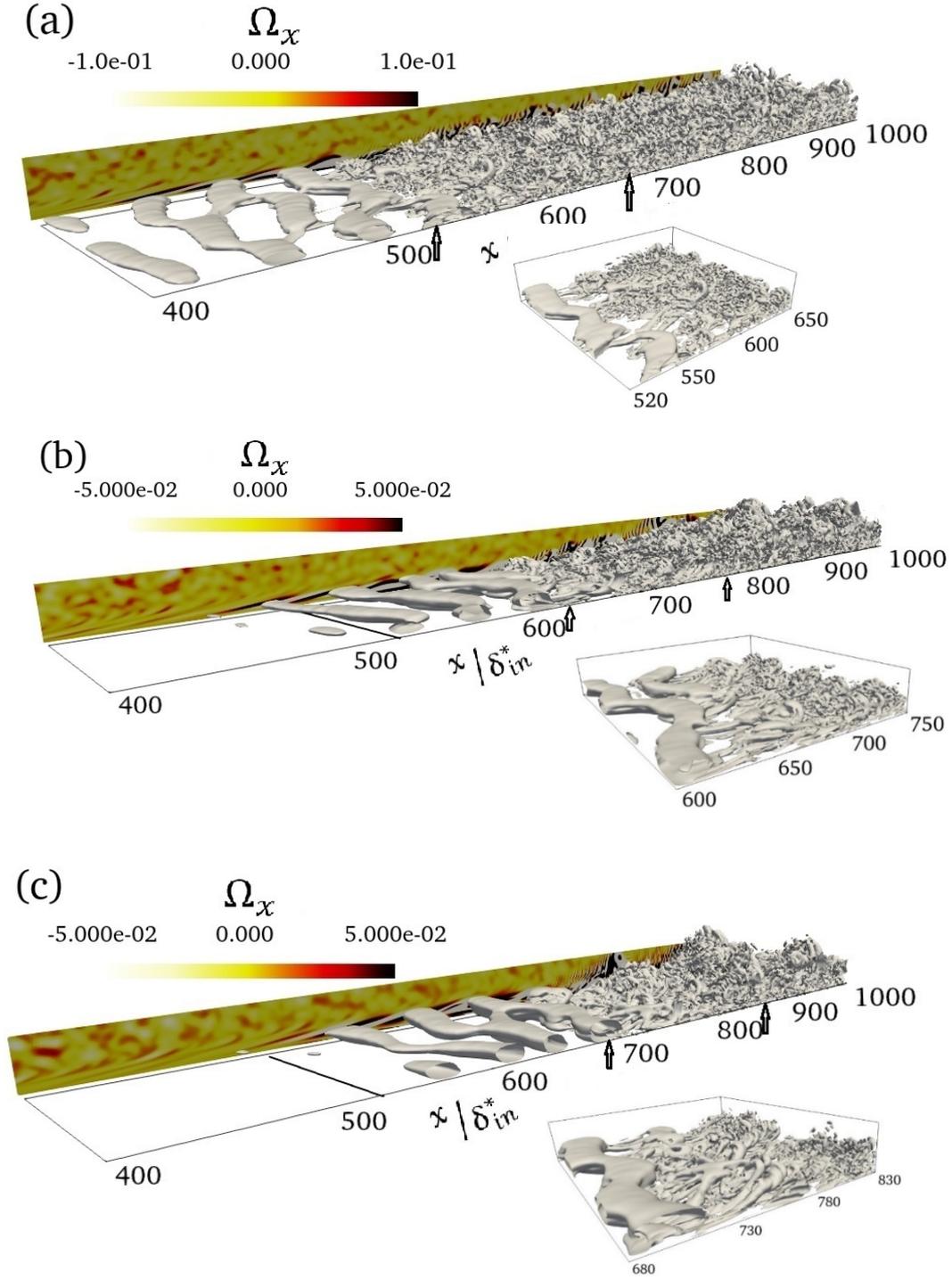

FIG. 9. Flow structures indicating the changes in instability characteristics with decreasing $Re_{\delta_{in}^*}$: (a) $Re_{\delta_{in}^*} = 158.7$; (b) $Re_{\delta_{in}^*} = 105.8$ and (c) $Re_{\delta_{in}^*} = 79.2$. The arrows in each subfigure indicate $x_{to}$ and $x_{te}$ respectively. The contours on the *x-y* plane correspond to the instantaneous vorticity $\Omega_x$. The iso-surfaces correspond to a particular value of $Q$: (a) $Q = 0.009$, (b) $Q = 0.005$ and (c) $Q = 0.003$.

The interaction between the streaks and instability waves seen in Figs. 7 and 8 is further illustrated by examining the flow structures evaluated using the $Q$ - criterion [11,12] applied to the instantaneous flow fields; see Fig. 9. The streaks are visualized in the $x$-$y$ plane at the mid-span location $\left(z/\delta_{in}^* = 90\right)$ using the contours of instantaneous streamwise vorticity ($\Omega_x$). For clarity, the plane is shifted to a side of the domain. The presence of negative and positive vorticity bands in the $\Omega_x$–contours near the wall indicates low- and high-speed streaks. For the $Q$ contours, values of $Q$ = 0.009, 0.005 and 0.003 are chosen respectively for $Re_{\delta_{in}^*}$ = 158.7, 105.8 and 79.2. The values of $Q$ chosen here are consistent with those used in Bose *et al.* [11]. The $Q$–isosurface plot reveals presence of spanwise rollers in the pre-transitional region for all the three cases. For the attached case, the rollers are considerably distorted just before the transition location (shown as an arow in Fig. 9(a)) due to the effect of streamwise streaks (Figs. 7(a), 8(a)). On the other hand, for the two separated cases, the rollers preserve their spanwise coherence (Fig 9) as the transition onset is approached, consistent with previous observations. The spanwise rollers break down soon after the onset of transition, resulting in Λ–shaped vortices which subsequently break down to generate smaller-scales of motion (Fig. 9).

To summarize, we have presented a unified picture of the changes in the stability characteristics for the attached and separated flow cases. The mixed-mode instability governing the three cases, consists of contributions from streamwise streaks and instability waves. There are two competing effects observed as we move from $Re_{\delta_{in}^*}$ = 158.7 (attached) to $Re_{\delta_{in}^*}$ = 79.2 (large separation). As the boundary layer separates, the inflection point moves away from the wall enhancing the inflectional instability (Figs. 4, 6, 9). At the same time, a decrease in the Reynolds number (from 158.7 to 79.2) pushes the location of onset of exponential growth downstream (Fig. 5) with distinct streamwise streaks visible over a longer streamwise extent for $Re_{\delta_{in}^*}$ = 79.2 (Figs. 7 & 8). As a result of these competing effects, for the attached boundary layer, the onset of transition is marked by an interaction between streaks and instability waves (wherein the spanwise rollers are modulated by the streaks). On the other hand, for the separated flow cases, the transition is dominated by the instability wave roll-up. What is interesting, though, is that the breakdown of the spanwise rollers at the start of transition zone shows common features among all the three cases (see insets in Fig. 9). The breakdown is seen to be driven by localised kinks in the rollers in spanwise direction (possibly due to a secondary instability; also see section V) and a rapid emergence of smaller scales of turbulence (although the severity of the breakdown reduces with decrease in $Re_{\delta_{in}^*}$; Fig. 9). These features are consistent with those reported in the previous literature [11, 24]. The fact that the "spanwise" breakdown of rollers is qualitatively similar for the attached and separated cases (despite differing contributions from the transient streak instability) seems to have an important bearing on the distribution of the "intermittency factor" within the transition zone for the three cases, which is the topic of the next section.

## IV. TURBULENT SPOTS AND INTERMITTENCY DISTRIBUTION

In this section, we analyse the time (space) localization of the high frequency (wavenumber) fluctuations in the time (space) series within the transition zone and determine transitional intermittency for the three cases. Figure 10 shows the evolution of the streamwise velocity fluctuations over a region including the location of onset of transition, i.e., the minimum in $C_f$ (Fig. 3). The time traces are obtained along the locus of local maximum in $\overline{u'^2}$, which is used as a representative location. Near the onset of transition, the time traces consist of relative low frequency fluctuations for the three cases (Fig. 10(a), (b) and (c)) which correspond to instability waves; for the attached case there could also be a contribution from the oscillation of the streamwise streaks (Figs 7(a) and 8(a)). Downstream of the transition location, the attached and separated cases exhibit a contrasting behaviour.

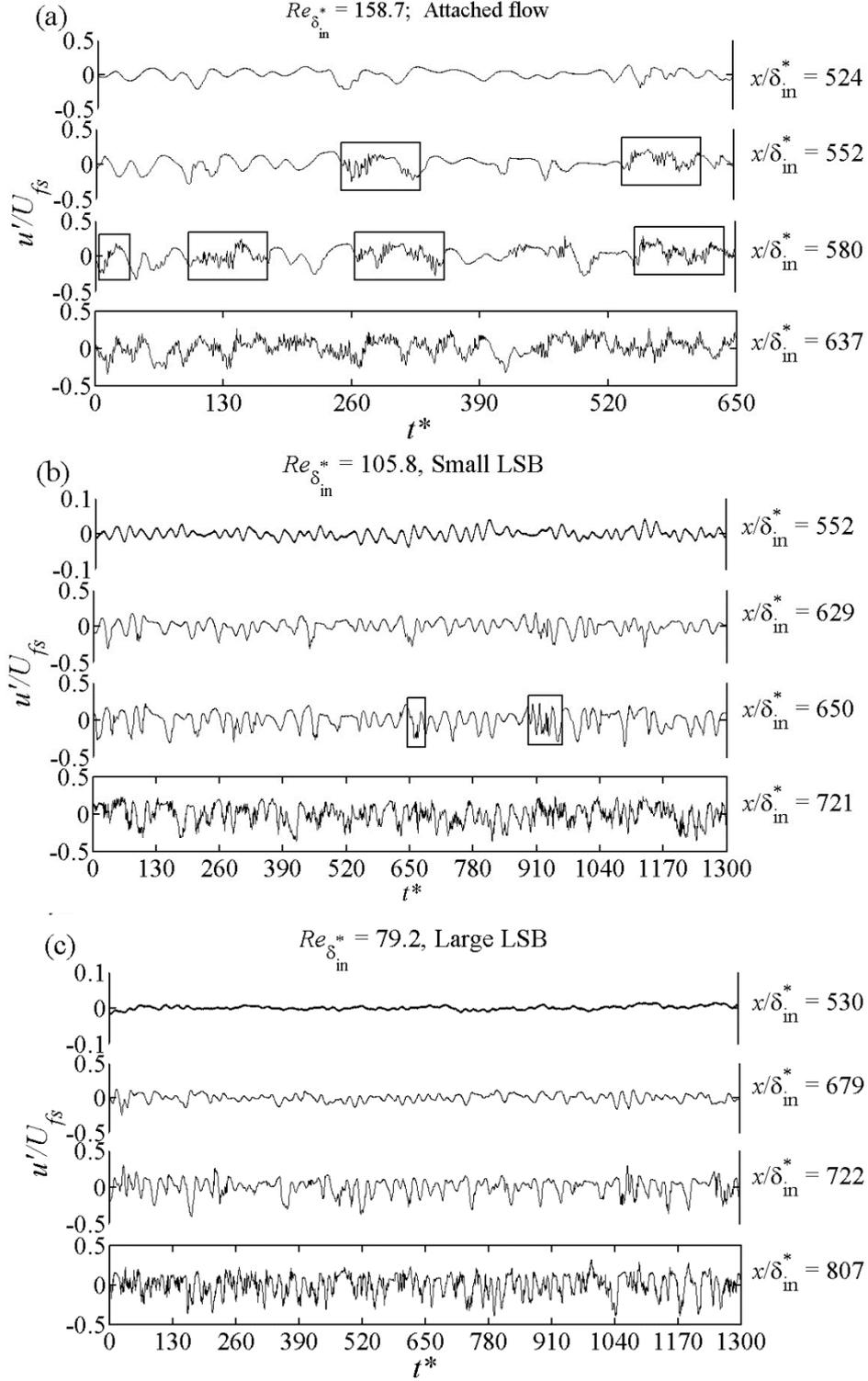

FIG. 10. Time traces of $u'$ at different streamwise locations along the locus of local maximum in $\overline{u'^2}$ (a) $Re_{\delta^*_{in}} = 158.7$ ($x_{to} = 524$, $x_{te} = 660$); (b) $Re_{\delta^*_{in}} = 105.8$ ($x_{to} = 620$, $x_{te} = 774$) and (c) $Re_{\delta^*_{in}} = 79.2$ ($x_{to} = 690$, $x_{te} = 857$). Here $t^* = \frac{tU_{fs}}{\delta^*}$.

For $Re_{\delta_{in}^*}$ = 158.7, one can clearly see the appearance of clusters of high frequency fluctuations localized in time, which indicate the presence of turbulent spots [41]. Here we follow the original definition of Emmons [38] in defining a turbulent spot i.e., a localized interval of high-frequency fluctuations in the velocity time series surrounded by quasi-laminar intervals. Note that the time traces from $x/\delta_{in}^*$ = 552 to 637 at this $Re_{\delta_{in}^*}$ (Fig. 10 (a)) are qualitatively similar to those obtained by Walker and Gostelow [13] for an attached boundary layer under a mild adverse pressure gradient. On the other extreme, for large separation at $Re_{\delta_{in}^*}$ = 79.2, the time traces post-transition do not indicate presence of distinct turbulent spots. Instead, the high frequency fluctuations are seen distributed over the entire time interval without any clear clustering; Fig. 10 (c). This is consistent with previous studies on separated flow transition [23-24, 42-43], which did not observe distinct turbulent spots in the separated shear layer (provided the separation bubble was sufficiently large). For the intermediate case of a small bubble ($Re_{\delta_{in}^*}$ = 105.8), the time traces show weak turbulent spots superposed on highly fluctuating velocity signal; see Fig 10 (b). The presence of turbulent spots in the velocity signal for a small separation bubble has been reported by McAuliffe and Yaras [44]. Overall, we see that as the Reynolds number decreases from 158.7 to 79.2, the velocity fluctuations within the transition zone become less clustered and more distributed over time (Fig. 10). This is consistent with the observations in Walker and Gostelow [13] that as an attached boundary layer approaches separation (say, through an increase in the strength of APG), the transitional velocity signal evolves continuously from (to use their terms) "random" behaviour, i.e., comprising of turbulent spots, to more "periodic" behaviour, i.e., not exhibiting distinct spots.

To further investigate the time localization behaviour of the transitional velocity signals for the three cases, we carry out a wavelet analysis of these signals. The wavelet transform is an effective technique to study the time-frequency behaviour of signals and has been previously used in the transition literature, e.g. [56-57]. Recently, Anand and Diwan [58] used this technique to contrast the "spotty" and "non-spotty" transition scenarios in an attached ZPG boundary layer downstream of a distributed roughness. In the present work, we follow the same style of presenting results from wavelet analysis as done in [58]. The velocity signals chosen for this analysis for the three cases are shown in Figs. 11 (a), (e) and (i); these correspond to the intermittency factor ($\gamma$) of ~ 0.6, which is approximately the middle of the transition zone. (The intermittency factor is defined as the fraction of the time a given velocity signal is turbulent; see Appendix A). Panels 11 (b), (f) and (j) present the contour plots of the pre-multiplied wavelet energy $\left(fC_w^2/U_{fs}^2\right)$, where $C_w$ is the wavelet coefficient amplitude. The pre-multiplied form is used as it allows identifying regions (on the log frequency axis) where the wavelet energy is primarily focused. Panels 11 (c), (g) and (k) show the pre-multiplied Fourier spectra, $f\phi_f$, where $\phi_f$ is the power spectral density. Note that time and frequency in Fig. 11 are scaled on the local displacement thickness ($\delta^*$) and local free-stream velocity $(U_{fs})$, $f^* = \frac{f\delta^*}{U_{fs}}$, $t^* = \frac{tU_{fs}}{\delta^*}$. The pre-multiplied Fourier spectrum shows a bi-modal shape for the attached flow case (Fig. 11 (c)) with the corresponding elevated energy levels seen in the wavelet contour plot (Fig. 11 (b)). For the two separated cases, the high-frequency hump (around $f^* = 0.2$) is not seen in the Fourier spectrum and the wavelet energy at these frequencies is seen to be lower than that for the attached case. The presence of bi-modal shape of the spectrum and the occurrence of distinct turbulent spots for the attached case are consistent with the similar observations in [58] for the spotty transition induced by distributed roughness. To bring out the time localization behaviour better we present high-pass filtered velocity signals in Fig. 11 (d), (h) and (l) with a cut-off frequency of $f^* = 0.1$. This frequency is chosen as it marks an effective partition of the fluctuation energy in the low and high frequency ranges (Fig. 11 (c), (g) and (k)). The fourth-order Butterworth filter is used towards this purpose. We have performed a sensitivity analysis with respect to the order of the digital filter and the cut-off frequency, and find that the qualitative features of the filtered signals are unaffected by these changes. For the attached flow case, there is a clear organisation of high frequency fluctuations in the form of turbulent spots separated by extended regions of laminar flow (Fig. 11 (d)). This organisation is also evident in the wavelet energy contours at high frequencies ($f^* > 0.1$) in Fig. 11 (b) (see [58]). For large separation ($Re_{\delta_{in}^*} = 79.2$), on the other hand, a clear organisation in terms of laminar patches and turbulent spots is absent, and high frequency fluctuations are more evenly distributed over the entire time interval (Fig. 11 (l)), except for some sporadic high frequency events seen in the wavelet energy contours for $f^* > 0.1$ (Fig. 11 (j)). This case is similar to the "uniform" transition scenario for a large bubble proposed by Wang [59] (wherein turbulent spots are not observed) which has similarities with the "periodic" transition reported in [13] as mentioned earlier. The small bubble in the present simulations ($Re_{\delta_{in}^*} = 105.8$) exhibits an intermediate behaviour to the two extreme cases. The filtered velocity signal for this case (Fig. 12 (h)) shows some organisation of high-frequencies suggesting presence of turbulent spots (see Fig. 10(b)) but the contrast between turbulent and non-turbulent intervals is less clear. In essence, the present simulation has captured the entire spectrum of transition scenarios, i.e., the "spotty" transition for $Re_{\delta_{in}^*}$ = 158.7 (attached case), "non-spotty" or "uniform" transition for $Re_{\delta_{in}^*}$ = 79.2 (large bubble) and an intermediate behaviour having both these features at $Re_{\delta_{in}^*}$ = 105.8 (small bubble).

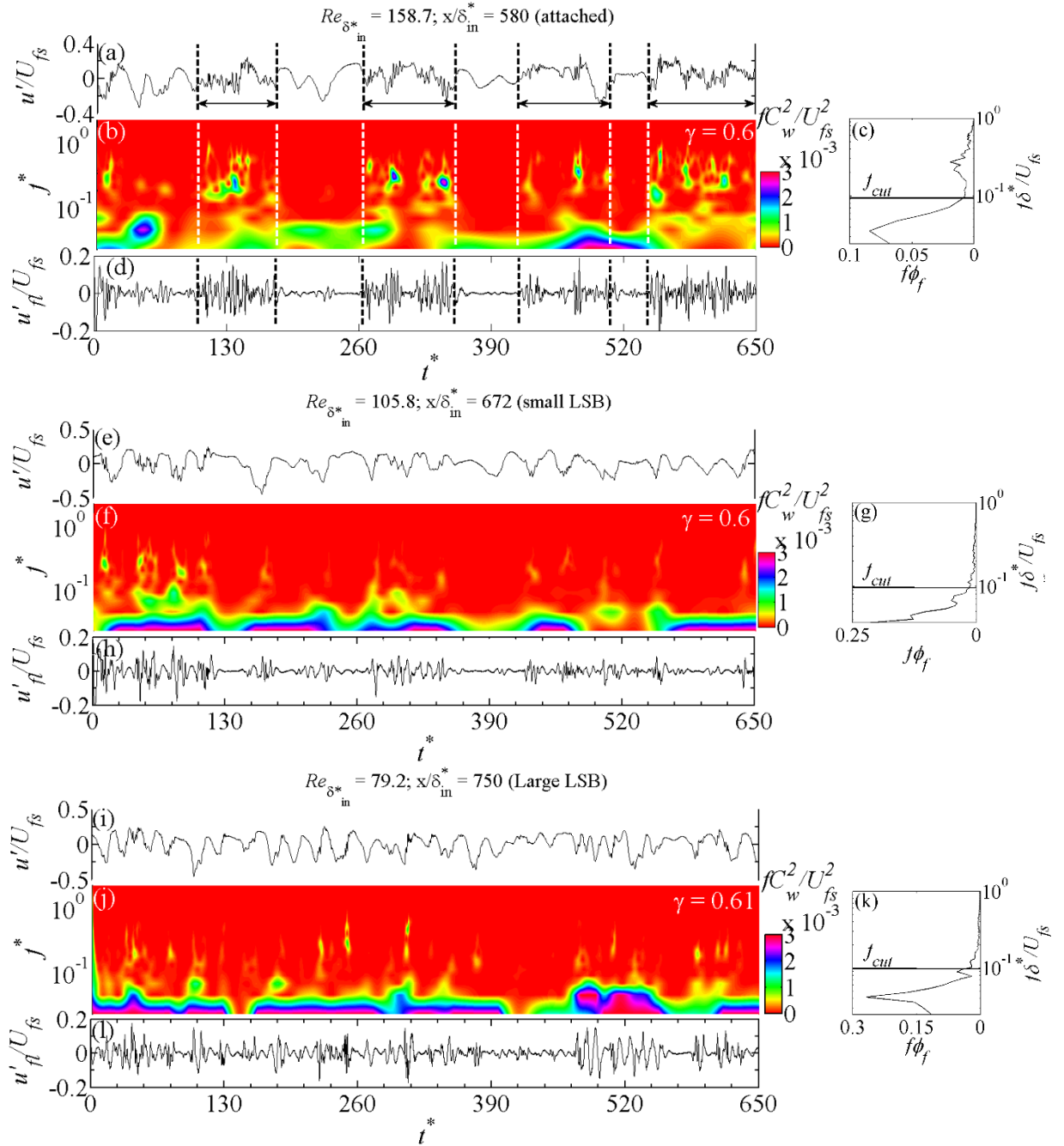

FIG. 11. Summary plot showing results from the wavelet and Fourier analysis for (a - d) $Re_{\delta_{in}^*}$ = 158.7; (e - h) $Re_{\delta_{in}^*}$ = 105.8 and (i - l) $Re_{\delta_{in}^*}$ = 79.2. The top, middle and bottom panels for each $Re_{\delta_{in}^*}$ represent, respectively, the original $u'$ time series over a fixed non-dimensional time interval (a, e, i), wavelet contour plot (b, f, j) and the filtered time series (d, h, l). The panels (c), (g) and (k) show the pre-multiplied Fourier spectra for the three $Re_{\delta_{in}^*}$.

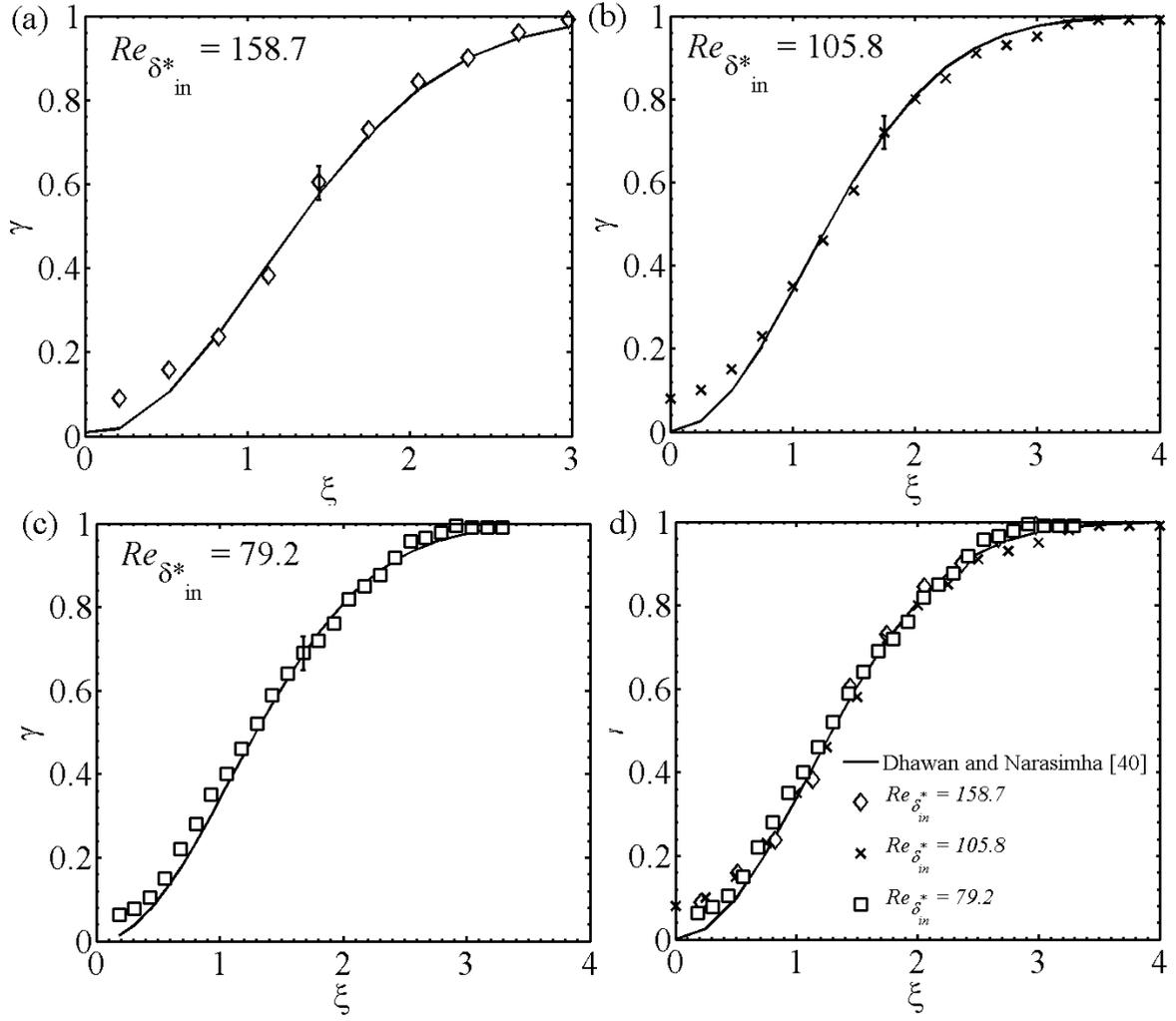

FIG 12. Distribution of intermittency factor and its comparison with universal the $\gamma$-curve of curve of Dhawan and Narasimha [40] (indicated by a solid line) for (a) $Re_{\delta^*_{in}}$ = 158.7; (b) $Re_{\delta^*_{in}}$ = 105.8 and (c) $Re_{\delta^*_{in}}$ = 79.2 and (d) all the three values of $Re_{\delta^*_{in}}$. The error bar corresponds to an uncertainty of ±3%; see Appendix A.

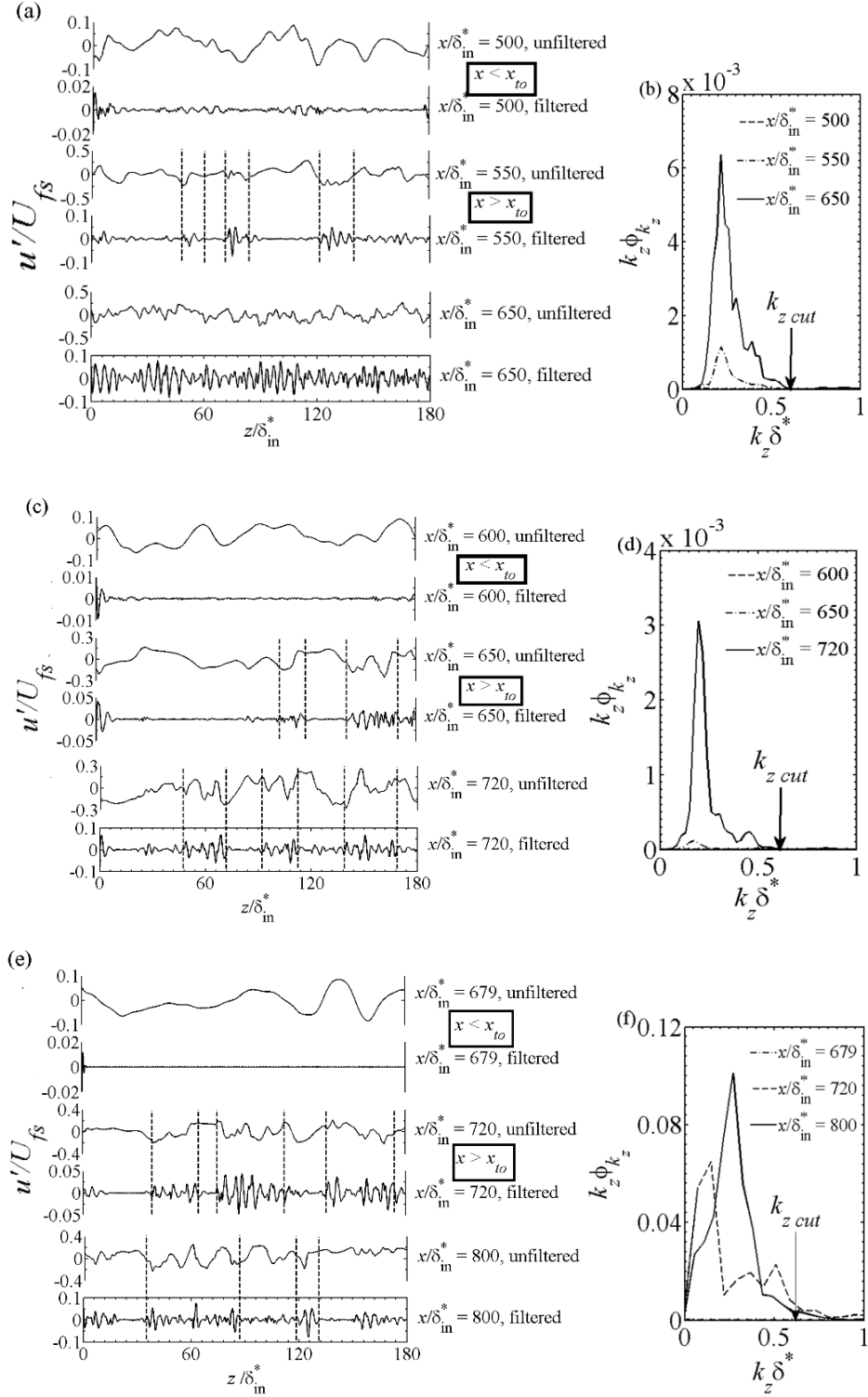

FIG. 13. Spanwise variation of streamwise velocity fluctuations illustrating concentrated regions of high wavenumber ($k_z$) events within the transition region for (a) $Re_{\delta_{in}^*} = 158.7$ ($x_{to} = 524, x_{te} = 660$); (c) $Re_{\delta_{in}^*} = 105.8$ ($x_{to} = 620, x_{te} = 774$) and (e) $Re_{\delta_{in}^*} = 79.2$ ($x_{to} = 690, x_{te} = 857$). The pre-multiplied wave-number spectra for the unfiltered velocity signals are shown in (b), (d) and (f).

Next, we plot the variation of the intermittency factor ($\gamma$), as a function of streamwise distance; see Fig. 12. Several methods have been reported in the literature for calculating $\gamma$ [60-61]. Here we prefer to use a version of the method by Hedley and Keffer [62] owing to its relative ease of implementation; this method has been used for calculating $\gamma$ in some of the recent studies [42-44, 58]. The calculated intermittency distributions are compared with the universal distribution of Dhawan and Narasimha [40]; see also Narasimha [39]. Following their work, the normalised streamwise distance is defined as $\xi = \frac{x-x_5}{x_{75}-x_{25}}$. Here $x_5$, $x_{25}$ and $x_{75}$ indicate the streamwise locations corresponding to 5%, 25% and 75% intermittency respectively. Figure 12 (a), (b) and (c) show that the universal distribution compares fairly well with the measured $\gamma$-distribution for all three cases, except for smaller values of $\xi$ where the calculated $\gamma$ values depart from the universal distribution. This can be attributed to the difficulties in accurately capturing low values of $\gamma$ using the method of Headly and Keffer [62]; such a limitation is also faced by others methods reported in the literature [60-61]. The kind of departure for low $\gamma$ seen in Fig. 12 has been observed before for attached APG and separated boundary layers [13,19, 43]. Fig. 12(d) compares the $\gamma$-distributions for the three cases, which are seen to compare well amongst themselves. More details on the intermittency calculation procedure are given in Appendix A, wherein we show "detector", "criterion" and "indicator" functions for two representative velocity signals – one each for the attached flow and large separation. We also present the effect of "smoothing period" on the value of $\gamma$ and use this exercise to estimate the uncertainty in $\gamma$ calculation to be $\pm 3\%$; this is shown in Fig. 12 as error bars. Since distinct turbulent spots are not observed for the large separation case, calculation of intermittency can pose problems as it is harder to detect the quiescent periods separating turbulent fluctuations (see Fig. A2 in Appendix A). However, we have used the Headley-Keffer method for this case also, so as to be consistent with the common practice in the literature of using standard intermittency calculation methods for analysing the "non-spotty" signals typical of separated flows.

Note that the universal $\gamma$–distribution in [39-40] was proposed for an attached boundary layer transition involving generation and propagation of turbulent spots. This condition is satisfied by the attached APG boundary layer in the present simulation ($Re_{\delta_{in}^*} = 158.7$) and therefore the favourable comparison with the universal $\gamma$-distribution for this case (Fig 12 (a)) is justifiable. This argument can also be extended for the small bubble ($Re_{\delta_{in}^*} = 105.8$) where the turbulent spots are still visible although less prominent (Fig. 10). It is, however, intuitively less clear why even for larger bubble ($Re_{\delta_{in}^*} = 79.2$), the $\gamma$-distribution should match well with the universal distribution, as in this case no distinct turbulent spots are seen. This observation has been reported in some of the previous studies [13, 19, 43, 63] and is known for some time. However, a satisfactory explanation towards this based on underlying physics has not been provided to the best of our knowledge. In this connection, we refer to the comment by Narasimha in response to a question regarding the validity of the universal $\gamma$–distribution for separation-bubble transition in the 2$^{nd}$ Minnowbrook workshop [45]. He pointed out that the universal distribution depends primarily on three postulates: the concentrated breakdown, a Poisson process for turbulence generation, and its linear propagation within the transition zone. The concentrated breakdown hypothesis has been originally stated as: "spots form at a preferred streamwise location randomly in time and cross-stream position" [41]. Referring back to Fig 9, we see that the onset of transition is fairly rapid for all the three cases; the spanwise rollers break down over a short distance from the location of transition onset to result in smaller-scale fluctuations (see the insets in Fig. 9). This suggests that the breakdown is concentrated (or at least limited to a short streamwise extent around the transition onset) for the attached as well as separated cases simulated herein. This is consistent with Vinod and Govindarajan [70]. The character of turbulent fluctuations as a function of time has been discussed so far – the turbulence generation is shown to be "spotty" for the attached case and "non-spotty" (or "uniform") for the large separation case (Fig. 11). However, it is also of interest to look at the behaviour of turbulent fluctuations in the spanwise direction, as it is relevant to the concentrated breakdown hypothesis. Note that this aspect has not much received attention in the literature.

Figure 13 presents the spanwise variation of $u'$ fluctuations for three streamwise locations for each case, at a suitably chosen time instant; the signals are chosen at the wall-normal location of the maximum in $\overline{u'^2}$. The power spectral density (in the premultiplied form) for these signals is plotted in Fig. 13 (b), (d) and (f) as a function of spanwise wavenumber $k_z$ (scaled on local $\delta^*$). As can be seen from the velocity signals for $x > x_{to}$, the energy in the higher wavenumbers goes on increasing for all the three cases. To clearly identify the nature of the high-wavenumber fluctuations, we apply the fourth-order Butterworth filter to the velocity signals in Fig. 13. The cut-off wavenumber chosen is $k_z \delta^* = 0.6$, which approximately marks the separation between the low- and high wavenumber fluctuations; Fig 13 (b), (d) and (f). The filtered velocity signals are presented in Fig. 13 (a), (c) and (e). It is evident that, just downstream of $x_{to}$, the high wavenumber fluctuations appear in clusters separated by quasi-laminar regions, akin to the turbulent spots appearing in time signals (Fig. 11). Interestingly, the clustering of high-wavenumber fluctuations is seen for all the three cases; Fig. 13 (a), (c) and (e). Thus, for the large separation case, even though the time signal does not show distinct turbulent spots (Fig. 11), the spanwise signals show a clear organisation or clustering typical of turbulent spots. That is, for this case, the time behaviour of turbulent fluctuations is more or less "uniform" and the spanwise behaviour is seen to be "spotty". On the other hand, for the attached case ($Re_{\delta_{in}^*}$ = 158.7), both the time and spanwise behaviour is "spotty" in nature (Fig. 11 and 13). It is therefore plausible that turbulent spots are generated for the large separation case also, although they do not distinctly appear in the time signals. The physical processes which might lead to this behaviour and its implication for the universal intermittency distribution are discussed in the next section.

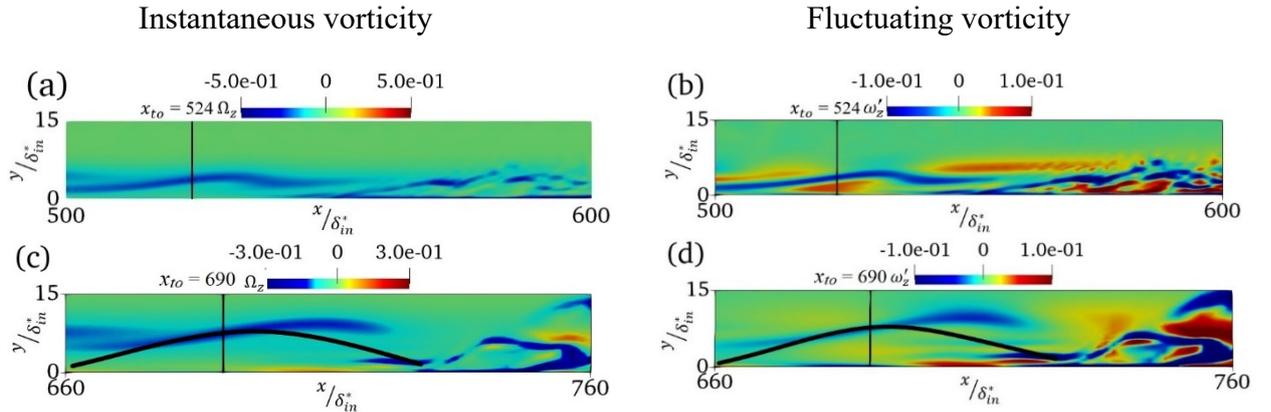

FIG. 14. Spanwise vorticity contours for (a), (b): attached case ($Re_{\delta_{in}^*}$ = 158.7) and (c), (d): large separation ($Re_{\delta_{in}^*}$ = 79.2). The left and right panels represent the instantaneous and fluctuating vorticity contours respectively.

## IV. PLAUSIBLE PHYSICAL CARTOON FOR THE TRANSITION ZONE

Here we provide a plausible mechanism for the transition processes based on the ideas of vortex-wall interaction and spanwise roller instability. Doligalski *et al.* [64] have shown that when a vortex is advected near a solid surface, an abrupt eruption of surface fluid takes place which leads to new vortex structures. They also point out that these eruptions take place at times (and positions) that cannot be easily predicted and that under such conditions the vortex motion becomes highly unsteady. Hatman and Wang [67] have applied these considerations to the separated flow transition. In the present context, it appears worthwhile applying these ideas to the spanwise rollers in Fig. 9, which can be considered as "vortical structures". For the attached case simulated here, the spanwise vortical structures are located close to the wall, as compared to the large bubble for which the structures are located in the separated shear layer away from the wall (Fig. 9). This is seen more clearly in the contours of the instantaneous spanwise vorticity shown in Fig. 14 (a) and (c). The near-wall vortical structures for the attached case can be expected to interact strongly with the wall and generate violent eruptions randomly in time, which would appear as distinct turbulent spots in the time signal (Fig. 10(a)). For the large bubble, however, the vortical structures are farther from the wall (compared to attached case; Fig. 14) and therefore, the vortex-wall interaction can be expected to be weaker in this case. The recirculation region, which is a region of weakened velocity gradients, can also act as a shield between the wall and vortical structures, further reducing the intensity of their interaction. As a result, for the larger bubble, the eruptions are likely to be less abrupt and localised, but more spread out (i.e., more "uniformly" distributed) in time; Fig. 10 (c). On the other hand, the spanwise breakdown of the rollers could be attributed to a centrifugal instability of vortical structures [68, 69], resulting in localised kinks that appear like "turbulent spots" in the spanwise signals (Fig. 13). Note that the mechanism can be expected to be present for all the cases, with the wall having a damping effect of differing degree for the three cases, with least damping for large separation. This is supported by the observation made earlier in relation to Fig. 9 (Section III.2) that the spanwise breakdown of rollers is qualitatively similar for the attached and separated cases. The combination of the above two factors could provide an explanation for the appearance of a "spotty" signal in time and spanwise directions for the attached case, and only in the spanwise direction for the large bubble case.

Based on the above discussion, we present a cartoon of the spot breakdown pattern at the location of transition onset, $x = x_{to}$; see Fig. 15. The cartoon in Fig. 15(a) corresponds to the standard transition scenario for the attached ZPG boundary layer, with turbulent spots appearing randomly in $t$ and $z$. The transition scenarios for attached APG boundary layer and large separation bubble are depicted in Fig. 15(b) and (c), which can be expected to exhibit a certain pattern of breakdown in the spanwise direction (with a characteristic wavelength $\lambda_p$) associated with the instability of the spanwise rollers as discussed above. Note that the spanwise rollers themselves are a consequence of the primary instability which is inflectional in nature (Fig. 4-6). For the ZPG boundary layer a pattern in the spanwise direction may be weak as distinct spanwise rollers are not exhibited by this flow, due to the fact that the primary instability is viscous in character [55]. These arguments are consistent with those in Vinod and Govindarajan [70], who found that the birth of turbulent spots was related to the pattern of instability in an APG boundary layer, whereas this connection was less clear for the ZPG boundary layer. Note that the breakdown of the spanwise rollers need not happen precisely at the same spanwise location for different time intervals and these locations can exhibit temporal jitter possibly introduced by the vortex-wall interaction alluded to earlier (Fig. 15(b) and (c)). (Incidentally this temporal jitter can be expected to introduce quasi-laminar patches in the velocity time series, thereby aiding in the calculation of intermittency factor for the separated flow cases; see Fig. 12 and Appendix A). For the large separation bubble, although the behaviour in time is non-spotty or "uniform" (Fig. 15(c)), it can be interpreted as the temporal spot generation rate being so high that a new spot is formed around the same location before the previous spot is advected downstream completely. The higher rate of spot generation for this case could be attributed to the weakened damping effect of wall on the breakdown of spanwise rollers, thereby enhancing the growth rates of the secondary roller instability. This could result in a "tailgating" in time of the successive spots, giving an appearance of a "uniform" time behaviour. Fig. 15(b) and (c) present deviations of the transition scenarios from the original form of the concentrated breakdown hypothesis [39] as depicted in Fig. 15(a); such deviations have been studied in the recent transition literature [70, 71]. Note that these scenarios are relevant for the processes near the location of onset of transition and it is not clear at this stage how the spots propagate in the downstream direction for the APG flows.

Based on the above discussion, we summarize the aspects of transition scenarios for the attached and separated flow cases relevant to the universal intermittency distribution.

- The postulate of concentrated breakdown is approximately valid for all the three cases.
- The spanwise traces of streamwise velocity fluctuations reveal distinct turbulent spots for all the three cases, with some spanwise regularity imposed by the breakdown of vortical rollers.
- The "uniform" appearance of turbulent fluctuations in time traces for large separation could be interpreted as tailgating of the turbulent spots generated at sufficiently high rate at the breakdown location. Thus, all the three cases exhibit spottiness in the transition zone with different manifestations.

Further, if we assume that the consecutive spots generated are independent of each other, it may be reasonable to expect that the spot generation is governed by the Poisson process. The above features can then be expected to approximately satisfy two of the three postulates [45] needed for the universal intermittency distribution to be valid, namely the concentrated breakdown and Poisson process for spot generation rate. We propose this to be a plausible explanation why the intermittency distribution for the large separation case shows a good comparison with the universal distribution, as seen in Fig. 12.

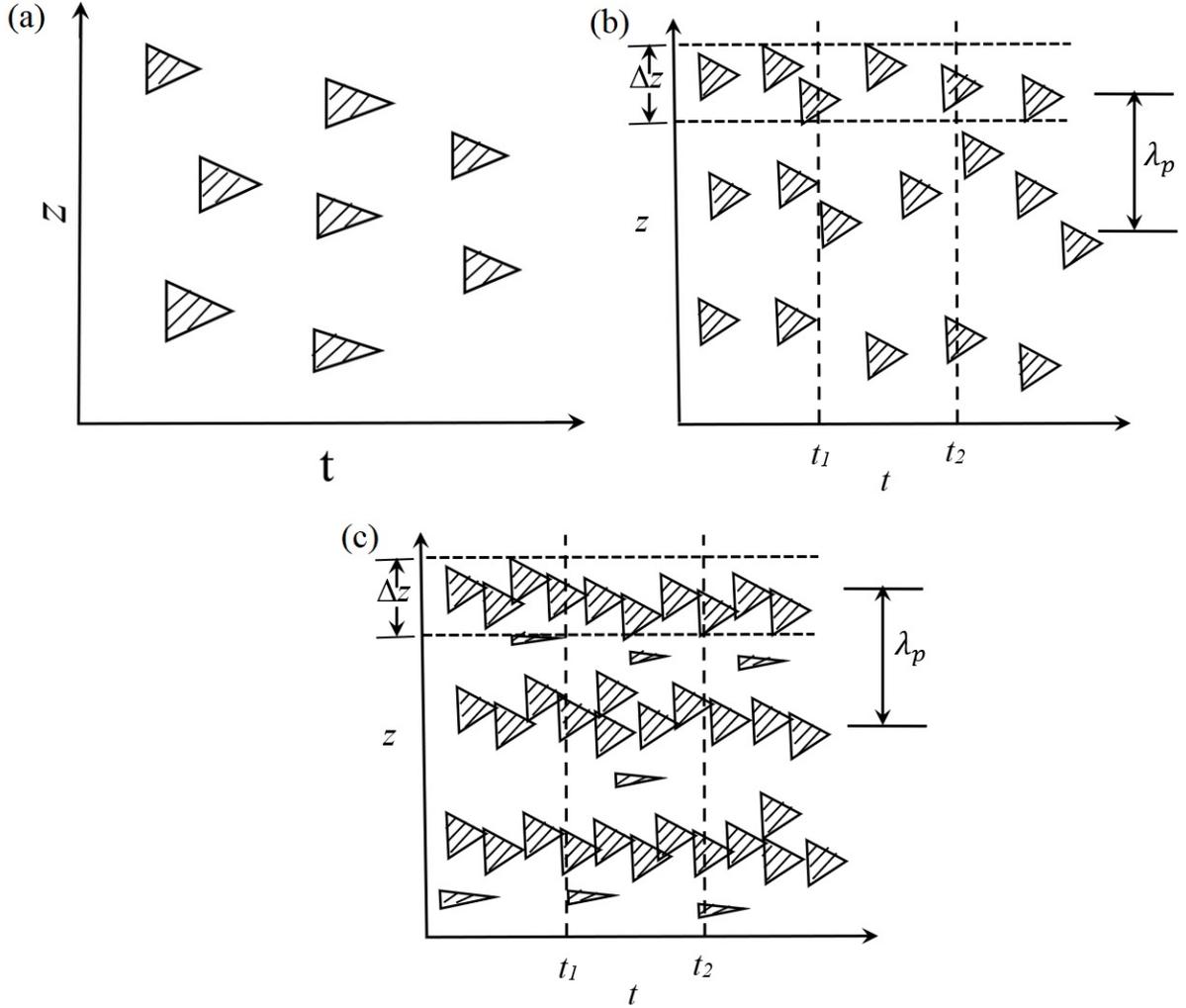

FIG. 15. Cartoon depicting occurrence of turbulence spots at the onset of transition ($x = x_{to}$) for (a): attached ZPG boundary layer, (b): attached APG boundary layer and (c): large separation bubble. The hatched regions indicate presence of high-frequency fluctuations. $\lambda_p$ is a typical wavelength of the pattern of spanwise breakdown of the rollers and $\Delta z$ is the temporal jitter associated with the breakdown locations. As a result of this jitter, the spot locations at two time-instants $t_1$ and $t_2$ will not be precisely identical. The small pointed triangles in Fig. 15(c) represent sporadic turbulent fluctuations that can appear randomly in time.

## VI. SUMMARY AND CONCLUSION

In this work we have reported DNS results on the transition in a boundary layer subjected to adverse pressure gradient, at a moderately low freestream turbulence intensity of 0.3%. This has enabled us to study, in a single setting, the changes in the instability mechanism and transition scenario as one moves from an attached boundary layer ($Re_{\delta_{in}^*}$ = 158.7) to a large separation bubble ($Re_{\delta_{in}^*}$ = 79.2), through the intermediate case of a small separation bubble ($Re_{\delta_{in}^*}$ = 105.8). We find, based on circumstantial evidence, that the instability mechanism for the three cases can be described as a mixed-mode instability involving contributions from the lift-up effect of streamwise streaks and the inflectional mechanism associated with the instability waves. A unified picture is presented of the changes in the stability characteristics as we move from attached to separated flow cases. For the attached case, the effect of streaks is felt right up to the location of transition onset, whereas for the separated cases streaks are much weaker near the transition onset. Notwithstanding this difference, the breakdown of spanwise vortical rollers at the beginning of the transition zone (that result in Λ–shaped vortices) is found to be qualitatively similar across the three cases.

To better understand the transition processes in the attached and separated cases we analyse the time-signals of streamwise velocity fluctuations using Fourier and wavelet transforms. For the attached flow case, the transitional time traces show a clustering of high-frequency fluctuations in the form of turbulent spots. On the other hand, for large separation bubble the time traces do not reveal distinct turbulent spots and high-frequency fluctuations appear more or less "uniformly" over the entire time interval. The case of small bubble shows features intermediate to the two extreme cases. The distribution of the intermittency factor within the transition zone is found to compare well with the universal intermittency–distribution of Narasimha [39], even for the large-separation case. We also analyse the "spanwise" signals of streamwise velocity fluctuations and show that they consist of clusters of high-wavenumber fluctuations separated by quasi-laminar regions, akin to turbulent spots, for attached as well as separated cases. Thus, for large separation the breakdown appears to be "spotty" in the spanwise direction but "non-spotty" in time. On the other hand, for the attached APG boundary layer both the time and spanwise velocity traces are spotty in character. We provide a plausible conjuncture for the physical processes that could lead to the observed behaviour of time and spanwise signals in different cases; a cartoon for the breakdown pattern of turbulent spots is also proposed. By interpreting the uniformly appearing turbulent fluctuations as tailgating among turbulent spots, we discuss plausible reasons why the intermittency distribution for large separation matches reasonably well with the Narasimha universal distribution.

## ACKNOWLEDGEMENT

We thank Prof O. N. Ramesh from the Department of Aerospace Engineering, IISc, Bengaluru for lending us the DNS code developed by Dr Saurabh Patwardhan. Thanks are due to Dr Abhijit Mitra for his help with numerical issues. SSD thanks IISc, Bengaluru for financial support through a start-up grant (No. 1205010620).

## AUTHOR'S CONTRIBUTIONS

All authors contributed equally to this work

## DATA AVAILABILITY STATEMENT

The data that support the findings of this study are available from the corresponding author upon reasonable request.

# APPENDIX A: DETAILS ON THE INTERMITTENCY CALCULATION METHOD

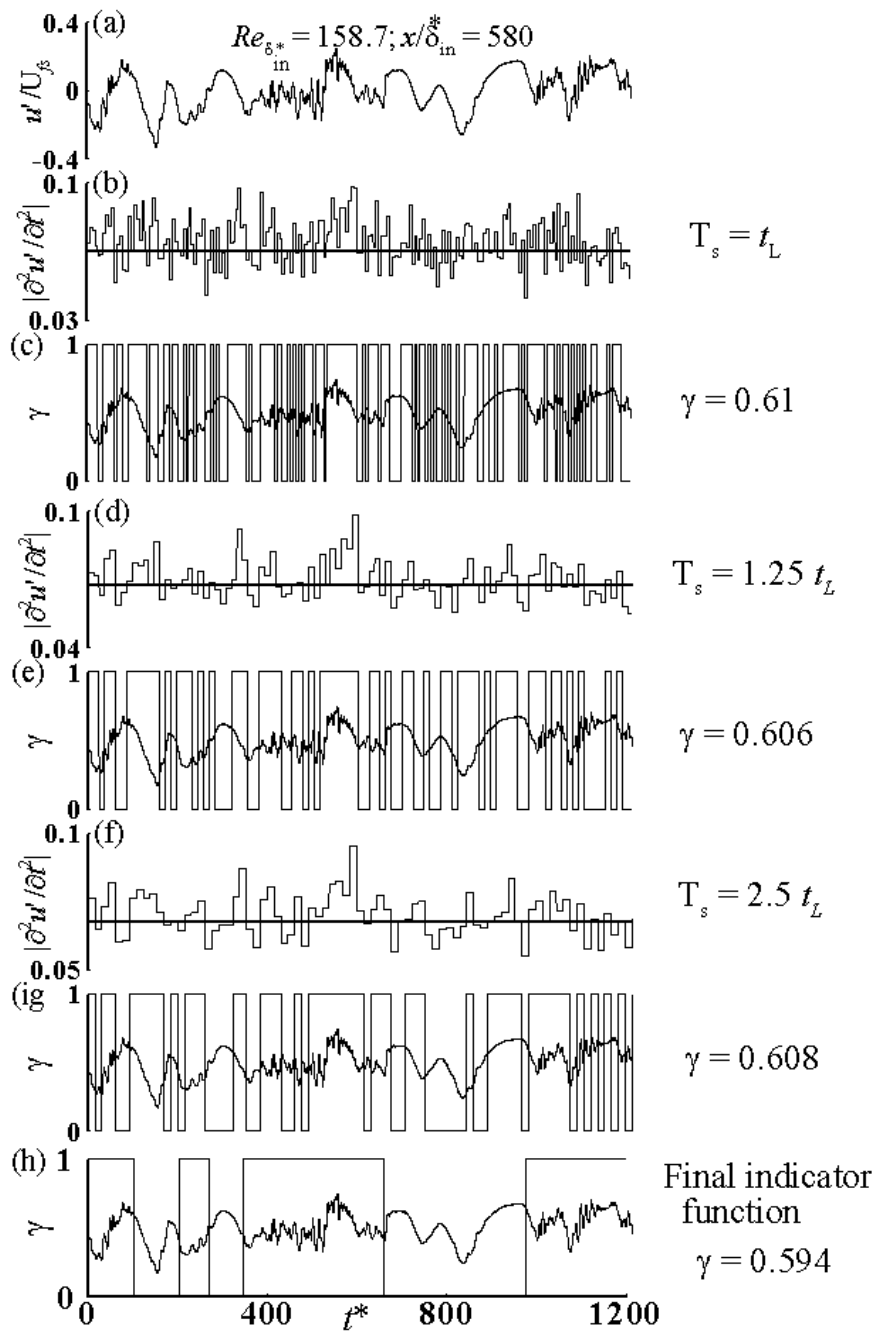

FIG. A1: Intermittency calculation exercise for varying smoothing periods for the attached boundary layer. (a) Velocity time trace. (b), (d), (f): Criterion functions for $T_s = t_L$, $1.25 t_L$ and $2.5 t_L$ respectively. (c), (e), (g): Indicator functions for $T_s = t_L$, $1.25 t_L$ and $2.5 t_L$ respectively superimposed on the time trace. (h) The final indicator function obtained by removing false zeros and ones by visual inspection. The $\gamma$ – values calculated for each case are shown beside the subplots. The horizontal lines in (b), (d) and (f) indicate the corresponding threshold levels.

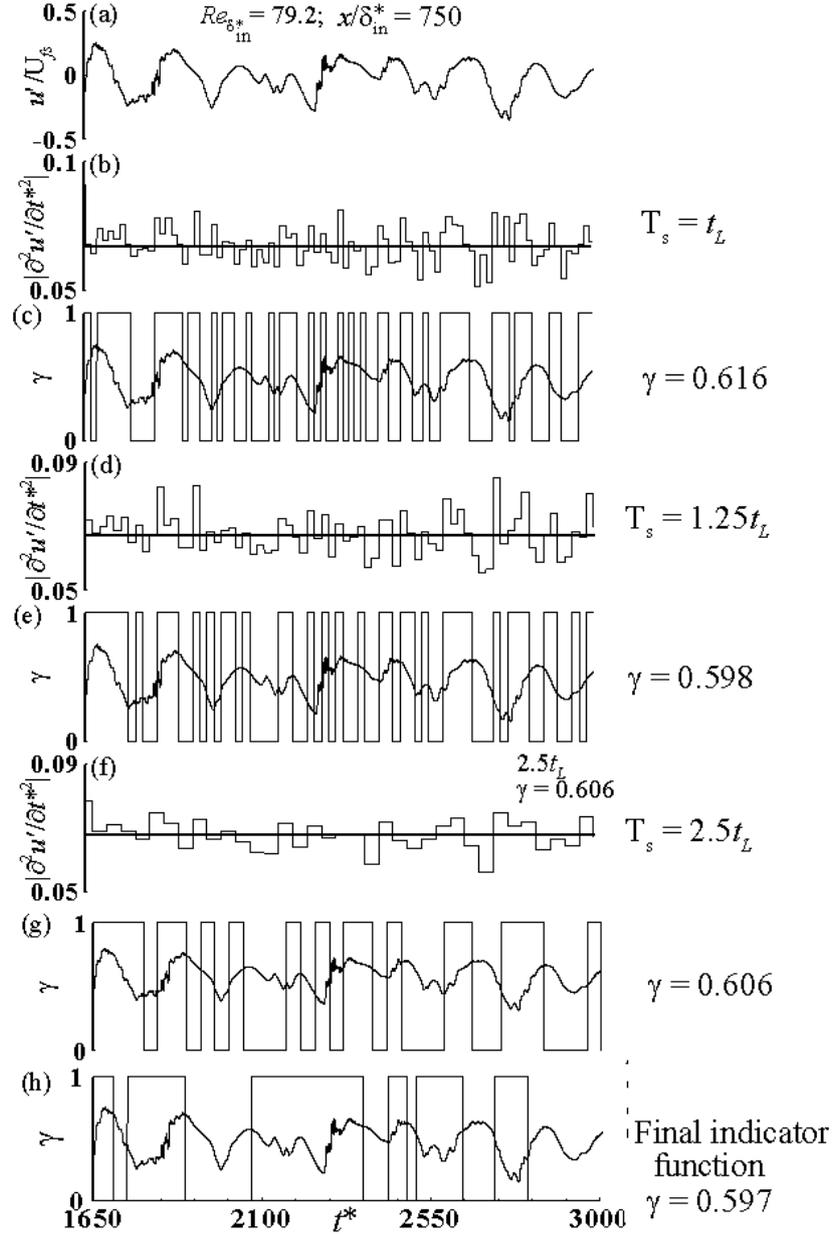

FIG. A2: Intermittency calculation exercise for varying smoothing periods for the large separation bubble. (a) Velocity time trace. (b), (d), (f): Criterion functions for $T_s = t_L$, $1.25t_L$ and $2.5t_L$ respectively. (c), (e), (g): Indicator functions for $T_s = t_L$, $1.25t_L$ and $2.5t_L$ respectively superimposed on the time trace. (h) The final indicator function obtained by removing false zeros and ones by visual inspection. The $\gamma$ – values calculated for each case are shown beside the subplots. The horizontal lines in (b), (d) and (f) indicate the corresponding threshold levels.

* Corresponding author: sdiwan@iisc.ac.in
[1] Current address: Department of Aerospace Engineering, Karunya Institute of Technology and Science, Coimbatore, Tamil Nadu, India – 641114. Email: ratnakumar@karunya.edu
[2] Current address: Associate Engineer, Caterpillar India Pvt Ltd, Bengaluru 560048. Email: karthikvnaicker@gmail.com

Here, we present some details on the intermittency calculation for the attached boundary layer ($Re_{\delta^*_{in}}$ = 158.7) and large separation bubble ($Re_{\delta^*_{in}}$ = 79.2), following the method proposed by Hedley and Keffer [62]. The time traces of the streamwise velocity fluctuations for attached and large separation cases are shown in Figs. A1 (a) and A2 (a) respectively. The time signals are sensitized using the modulus of the second derivative and are termed as "detector" functions [62]. The intermittency calculation following a cut-off on the detector function usually results in false zeros and ones [62] in the "indicator" function which need to be estimated. This is typically done by smoothing the detector function using short-time averaging, which results in the "criterion" function [62]. Headley and Keffer [62] suggested the smoothing time period (T$_s$) to be equal to 4$\Delta T$ for a ZPG boundary layer, where $\Delta T$ is the sampling period of the velocity signals. Different smoothing periods have been reported by other investigators [43, 73, 74]. In the present work, we find that 4$\Delta T$ is not a sufficiently long smoothing period to get a reasonably good criterion function and prefer to use the large-eddy turnover time $\left(t_L = \delta/U_{fs}\right)$ to be a more convenient estimate for T$_s$. Figures A1 (b) and A2 (b) show the criterion functions respectively for attached case and large separation, with T$_s$ = $t_L$. A cut-off on the criterion function is obtained from the region of maximum curvature of the cumulative probability distribution [62] and shown as a horizontal black line in Figs. A1 (b) and A2 (b) (and also in other criterion function plots). The resulting indicator function representing the intermittency variation superimposed on the velocity time traces are plotted in Figs. A1 (c) and A2 (c). As can be seen, one can still detect the presence of false zeros and ones in the indicator functions. Towards this, we have used two higher values of T$_s$ = 1.25$t_L$ and T$_s$ = 2.5$t_L$, which result in criterion functions shown in Figs. A1 (d, f) and A2 (d, f); T$_s$ = 2.5$t_L$ is found to give a much more satisfactory match between the time trace and indicator function for both the attached and separated cases (Fig. A1 (g) and A2 (g)). However, there are still some false zeros and ones present in the indicator function and they have been removed by means of visual inspection; the resulting functions are shown in Figs. A1 (h) and A2 (h) for the two cases. The visual inspection method although somewhat subjective, has been found in the literature to be generally useful in obtaining accurate estimates of the intermittency factor [72-74]. It must be noted that, for the attached boundary layer, the turbulent spots are interspersed by laminar regions and hence can be identified with relative ease (Fig. A1(h)). For large separation, however, distinct turbulent spots are absent and turbulence appears more "uniformly" in time (Fig. A2 (h)). This can present difficulties in removing false zeros and ones, which again is a well-recognized fact in the literature [13, 43, 72-74].

The $\gamma$–distributions presented in Fig. 12 have been obtained by using $T_s = 2.5 t_L$ and removing the false zeros and ones by visual inspection. The present exercise of varying the smoothing period and using the visual inspection method give the total variability of the calculated $\gamma$ values (presented in Figs. A1 & A2 for each $T_s$) to be about 3%. We therefore consider the uncertainty in calculation of $\gamma$ to be $\pm 3\%$, which is indicated by error bars on the $\gamma$–distributions in Fig. 12.